\documentclass[fleqn,usenatbib]{mnras}

\usepackage{mathptmx}

\usepackage[T1]{fontenc}

\DeclareRobustCommand{\VAN}[3]{#2}
\let\VANthebibliography\thebibliography
\def\thebibliography{\DeclareRobustCommand{\VAN}[3]{##3}\VANthebibliography}

\usepackage{graphicx}	
\usepackage{amsmath}	
\usepackage{amssymb}	
\usepackage[dvipsnames]{xcolor}
\usepackage{tikz}
\usetikzlibrary{calc,patterns,decorations.pathmorphing,decorations.markings}
\usetikzlibrary{quotes,angles}

\DeclareMathOperator{\sgn}{sgn}

\title[]{Equivalence between simple multilayered and homogeneous laboratory-based rheological models in planetary science}

\author[Y. Gevorgyan et al.]{
Yeva Gevorgyan,$^{1,2}$\thanks{E-mail: yeva@ime.usp.br (YG)}
Isamu Matsuyama,$^{1}$
and Clodoaldo Ragazzo$^{2}$
\\
$^{1}$Lunar and Planetary Laboratory, University of Arizona, Tucson, AZ 85721, USA\\
$^{2}$Instituto de Matem\'{a}tica e Estat\'{i}stica, Universidade de S\~{a}o Paulo, 05508-090 S\~{a}o Paulo, SP, Brazil
}

\date{Accepted 2023 May 14. Received 2023 April 17; in original form 2023 March 07}

\pubyear{2023}

\begin{document}
\label{firstpage}
\pagerange{\pageref{firstpage}--\pageref{lastpage}}
\maketitle

\begin{abstract}
The goal of this work is to investigate under which circumstances the tidal response of a stratified body can be approximated by that of a homogeneous body. We show that any multilayered planet model can be approximated by a homogeneous body, with the same dissipation of tidal energy as a function of the excitation frequency, as long as the rheology of the homogeneous model is sufficiently complex. Moreover we provide two straightforward methods for finding the parameters of the homogeneous rheology that would exhibit the same tidal response as the layered body. These results highlight the fact that the two models cannot be distinguished from each other only by the measurement of the second degree tidal Love number and quality factor, and that we do not need the complexity of the multilayer planet model in order to estimate its tidal dissipation. The methodology promises a great simplification of the treatment of multilayered bodies in numerical simulations because the treatment of a homogeneous body---even with a complex rheological model---can be computationally better handled than that of a multilayered planet.

\end{abstract}




\section{Introduction}

Since the discovery of the first exoplanet by \cite{Mayor1995} over 5000 exoplanets and almost 4000 exoplanetary systems have been identified and cataloged, hence it is becoming necessary to consider tidal interactions in exoplanetary systems. The tidal response of a planet depends on the timescale of the perturbation and on the rheological properties of its interior (i.e., elasticity, density, and viscosity) \citep{Bagheri2022}. Many moons and planets within the solar system exhibit a stratified internal structure, as confirmed by seismic and moment of inertia data. For the Earth, for example, the multilayered Preliminary Reference Earth model (PREM) is widely accepted \citep{Dziewonski1981}. Bodies like the Moon \citep{Harada2014} and Mercury \citep{Goossens2022} are also known to have a stratified structure. Some icy moons have clear evidence of subsurface oceans and consequently stratified internal structure, e.g. Enceladus \citep{THOMAS201637}, Europa \citep{Carr1998}, Titan \citep{iess2012}. Several models that account for the layered internal structure of these bodies were proposed \citep{Matsuyama2014, Matsuyama2018, Folonier2017,Boue2017,Bolmont2020,ragazzo2022librations}. The stratified internal structure of the Solar System bodies suggests complex internal structure for the extrasolar planets as well. While the observational data in our Solar System is relatively abundant, we have few observational constraints for extrasolar planets. Stratified models are usually complex and require to fit numerous parameters from the observational data. Models with few adjustable parameters can be a good alternative to study exoplanets with scarcity of data, or to investigate rotational and orbital evolution of nearby objects, since considering a multilayered structure does not significantly impact the rotational states of the planets \citep{Walterova2017}. One possible approach is to model the otherwise stratified body after an effective homogeneous rheological model \citep{RaR2017,gev2020,Gev2021}. The rheological models used are usually the ones proposed based on visco-elastic behavior of planetary materials in laboratory conditions, such as Maxwell, Voigt, Burgers, Andrade \citep{Andrade1910}, Sundberg-Cooper \citep{Sundberg2010} or their combinations. The natural question is to what extent stratified bodies can be modeled with homogeneous rheological models. 

The use of a homogeneous body to approximate the tidal response of a multilayered structured body, was recently addressed in \citet{Bolmont2020}. The authors conclude that it is possible to approximate the response of a multilayered planet by that of a homogeneous planet for purely rocky bodies, but not for planets with both rocky and icy layers due to the large differences in material properties (in particular the viscosity), which results in additional peaks in the frequency dependence of tidal dissipation. However, \citet{Gev2021} show that the tidal response of a planet with large viscosity variations between layers can in fact be approximated by a homogeneous body, if the rheology of the homogeneous model is slightly more complex than the one used in \citet{Bolmont2020}. 


In this work, we revisit the problem of approximating the tidal response of a stratified body by that of a homogeneous body.  We show that any multilayered body can be approximated by a homogeneous body, with the same dissipation of tidal energy as a function of the excitation frequency, as long as the rheology of the homogeneous model is sufficiently complex. 

The paper is structured as follows: In Section \ref{Dissipation}, we shortly discuss the dissipation models used in the paper. In Section \ref{Homogeneous}, we propose a way to associate a homogeneous rheology to a given stratified celestial object. 
In Section \ref{Simplified}, we provide a recipe to simplify the homogeneous rheology preserving the rheological behavior in an interval of frequencies of interest. In a second step we show that from the homogeneous model with complex rheology we may extract a "minimal model" which is much simpler but still reproduces the rheological behavior of the multilayered body with accuracy. 
Finally, In the Section \ref{conclusions}, we summarize the main results and discuss the implications of our findings.
\section{Dissipation models}\label{Dissipation}

We consider a deformable body orbiting a point-like perturbing body and experiencing tidal deformations under its influence. Tidal deformation of a celestial body results in both vertical and horizontal displacement of its surface and in the ensuing perturbation of its gravitational field, which is described in terms of Love numbers. The Love numbers depend on the timescale of the perturbation and on the rheological properties of the interior (i.e., elasticity, density, and viscosity). In particular, the degree-2 tidal Love number $k_2$ \citep{Love1911} quantifies the ability of a celestial body to respond to degree-2 tidal forcing. For a perfectly elastic body the deformation is instantaneous, and the tidal bulges are aligned with the direction of the perturbing body. There is no tidal evolution in this case. However, for a real body, the response is never perfectly elastic and part of the response is dissipative, resulting in a delay or lag in the deformation. Here we are interested in comparing the dissipative behavior of a body modeled assuming a homogeneous or stratified interior structure.

\cite{gev2020} obtained an analytic formula for the tidal energy dissipation rate of a body in $1{:}1$ spin-orbit resonance on a slightly eccentric planar orbit in the presence of forced librations. Ignoring forced libration contributions (here we are not interested in this contribution, which could be added to the problem if needed), 
we obtain the commonly used expression for 
the time- and volume-averaged tidal energy 
dissipation rate (e.g., \cite{Poirier.1983}, Eq. 24; \cite{Segatz1988}, Eq. 13)
\begin{equation}\label{disen}
\frac{\Delta E}{T}=\; - \Im\left[k_{2}(i\omega)\right]\frac{(nR)^5}{G}\left(\frac{m_2}{m_1+m_2}\right)^2\times\frac{21}{2}e^2,
\end{equation}
where $k_{2}(i\omega)$, $\omega$, $n$, $R$, $G$, $m_1$, $m_2$, and $e$ are, respectively, the $2$nd-degree tidal Love number, tidal forcing frequency, mean rotation rate, mean radius, gravitational constant, mass of the deformable body, mass of the perturbing body, and orbit eccentricity. The frequency-dependent part of the dissipation rate is encoded in the imaginary part of the tidal Love number, $k_{2}(i\omega)$, hence we will use it instead of the dissipation rate in the paper.

\subsection{Dissipation in a stratified body}\label{stratified}
The tidal response of a stratified body and the corresponding Love numbers can be found by solving the mass conservation, momentum conservation, and Poisson equations. We use the classical propagator matrix method to solve these equations \citep[e.g.][]{sabadini2016global}. 
The solid core is treated as in \citet[Appendix A]{Matsuyama2018}. Each layer is assumed to be either solid with linear viscoelastic rheology or liquid and inviscid.
Viscoelastic layers are assumed to have a Maxwell rheology characterized by an elastic shear modulus and a viscosity.
Liquid layers can be treated using the method of \citet{jara2011effects}, or by assuming a viscoelastic layer with shear modulus approaching zero, which yields virtually the same results. 

\begin{figure}
\begin{center}
\begin{tikzpicture}[scale=0.93, transform shape]
\node[inner sep=0pt] (1) at (0,0)
{\includegraphics[width=.312\textwidth]{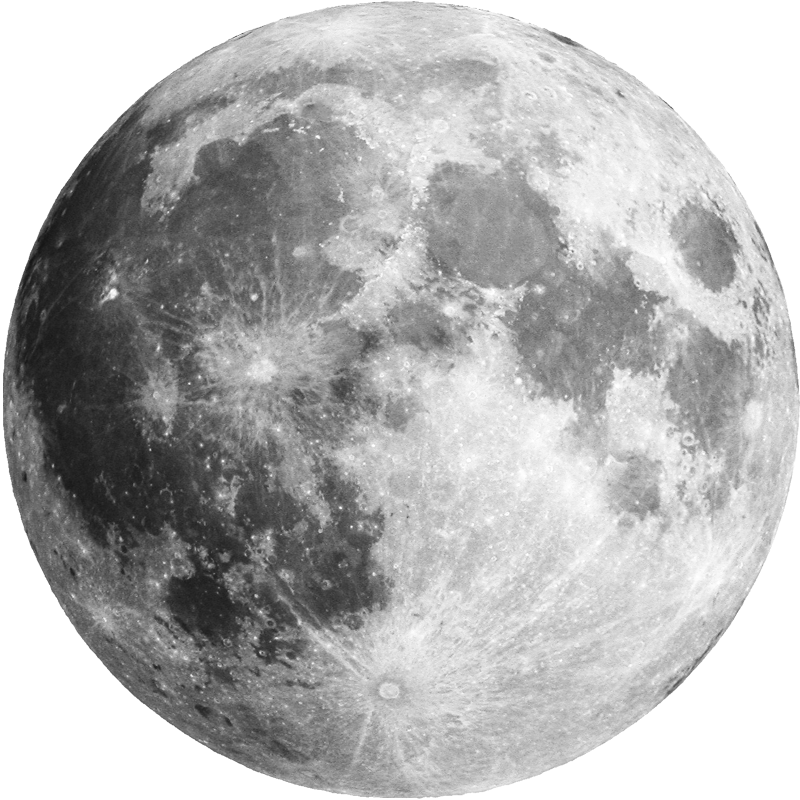}};
\draw[fill=orange!90] (-90:0) coordinate (beta) arc (-90:90:0) coordinate (alpha) -- (90:0.368) arc (90:-90:0.368) -- cycle ;
\draw[fill=yellow!90] (-90:0.368) coordinate (beta) arc (-90:90:0.368) coordinate (alpha) -- (90:0.517) arc (90:-90:0.517) -- cycle ;
\draw[fill=Apricot!60] (-90:0.517) coordinate (beta) arc (-90:90:0.517) coordinate (alpha) -- (90:0.8) arc (90:-90:0.8) -- cycle ;
\draw[fill=OliveGreen!70] (-90:0.8) coordinate (beta) arc (-90:90:0.8) coordinate (alpha) -- (90:2.7) arc (90:-90:2.7) -- cycle ;
\coordinate (o) at (0,0);

            \draw [very thin] (-3,0) -- (0,0);
            \node at (-5,0) {Solid core};
            \draw [very thin] (-3,0.465) -- (0,0.465);
            \node at (-5,0.465) {Liquid core};   
            \draw [very thin] (-3,-0.6265) -- (0,-0.6265);
            \node at (-5,-0.6265) {Low viscosity layer};  
            \draw [very thin] (-3,1.66) -- (0,1.66);
            \node at (-5,1.66) {Mantle};

            \draw [very thin] (-3,2.732) -- (0,2.732);
            \node at (-5,2.732) {Crust};            

\draw[very thin] (alpha)--(o)--(beta);
\end{tikzpicture}
\end{center}
\caption{Moon internal structure.}
\label{moon_int}
\end{figure}
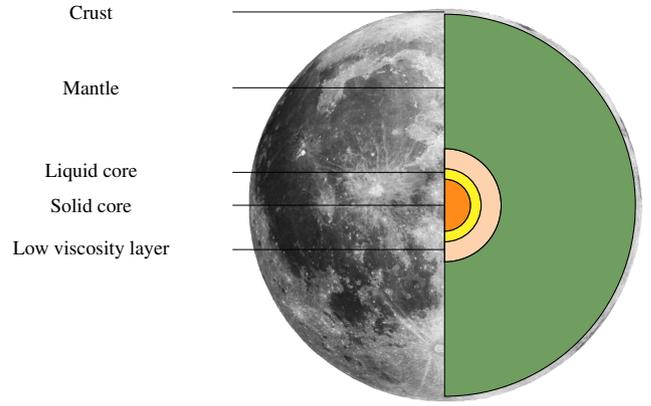

\begin{table*}
\begin{center}
\caption{Lunar interior structure parameters.}

\begin{tabular}{c c c c c}
\hline
The layer \hspace{1.0cm} & Outer radius ($\mathrm{km}$) \hspace{1.0cm} & Density ($\mathrm{kg/ m^3}$) \hspace{1.0cm} & Rigidity ($\mathrm{GPa}$) & Viscosity ($\mathrm{Pa s}$) \\
\hline
Solid core \hspace{1.0cm} & $213.517$ \hspace{1.0cm} & $7720.16$ \hspace{1.0cm} & $40$ & $10^{21}$ \\
Liquid core \hspace{1.0cm} & $325$ \hspace{1.0cm} & $6700$ \hspace{1.0cm} & $10^{-10}$ & $1$ \\
Low viscosity layer \hspace{1.0cm} & $500$ \hspace{1.0cm} & $3800$ \hspace{1.0cm} & $60$ & $5\times10^{16}$ \\
Mantle \hspace{1.0cm} & $1697.15$ \hspace{1.0cm} & $3356$ \hspace{1.0cm} & $62.5$ & $10^{21}$ \\
Crust \hspace{1.0cm} & $1737.15$ \hspace{1.0cm} & $2735$ \hspace{1.0cm} & $15$ & $10^{23}$ \\
\hline
\end{tabular} \\[0.3em]
{\footnotesize 
We assume a 5-layer interior structure based on the mass, moment of inertia, and gravity constraints \citep[][Table 2]{Matsuyama2016}. The core density is adjusted to satisfy the mass and moment of inertia constraints. All parameters are within the uncertainties in \citet{Matsuyama2016}. The core and mantle viscosity of $10^{21}$ Pa s is representative of Earth's mantle, and the crust viscosity of $10^{23}$ Pa s is representative of Earth's lithosphere.}
\label{moon_profile}
\end{center}
\end{table*}

We consider the specific case of the Moon
to illustrate the theoretical results below, assuming a 5-layer interior structure in Figure \ref{moon_int} consisting of a solid core, liquid core, low viscosity layer, mantle and crust, based on mass, moment of inertia, and gravity constraints \citep{Matsuyama2016}, as summarized in Table \ref{moon_profile}.
The solid core radius and density were adjusted to satisfy to mass and moment of inertia constraints. 
The presence of a low viscosity layer at the base of the mantle is consistent with the seismic constraints obtained by 
\citet{Weber.2011}; 
however, the seismic model of \citet{Garcia.2011} does not include such a layer.
\citet{Harada2014} illustrated that a low viscosity layer at the base of the mantle can explain the frequency dependence of the tidal quality factor $Q$\footnote{The quality factor is defined as minus the ratio between  the real and the imaginary parts of the Love number.} \citep{Williams2015} assuming a simple Maxwell rheology. \citet{Nimmo2012} attribute the observed frequency dependence to an absorption band due to grain boundary sliding. Although authors were able to fit the monthly $Q$ and $k_2$, the model that they used to obtained the correct sign of the $Q$'s slope did not match $k_2$. Recently, \citet{Walterova2023} concluded that the available selenodetic tidal parameters are insufficient to distinguish a weak basal layer above the lunar core from the manifestation of elastically accommodated GBS in the mante.

\begin{figure}
\centering
\includegraphics[scale=0.46]{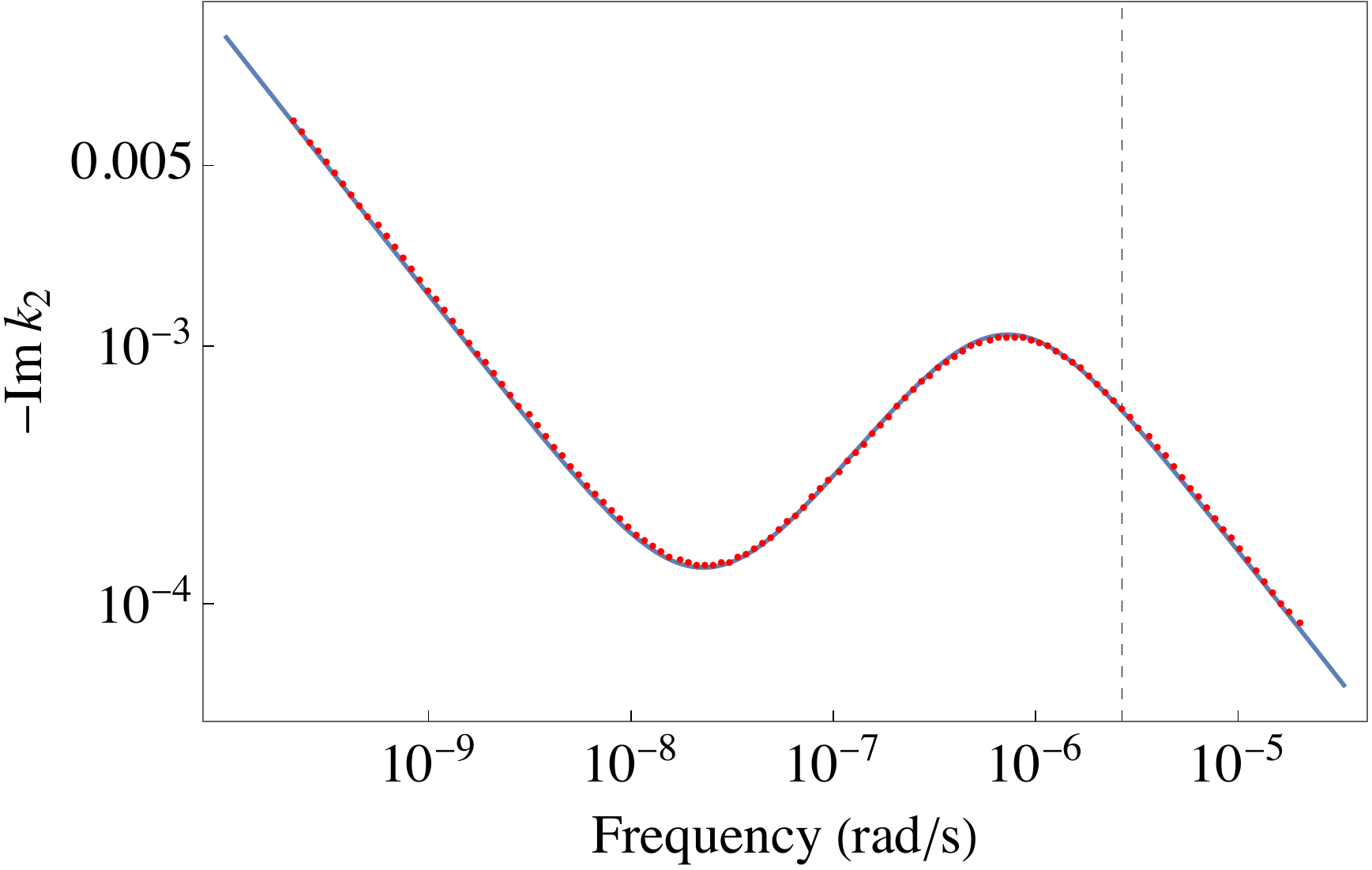}\\
\vspace{2em}
\includegraphics[scale=0.46]{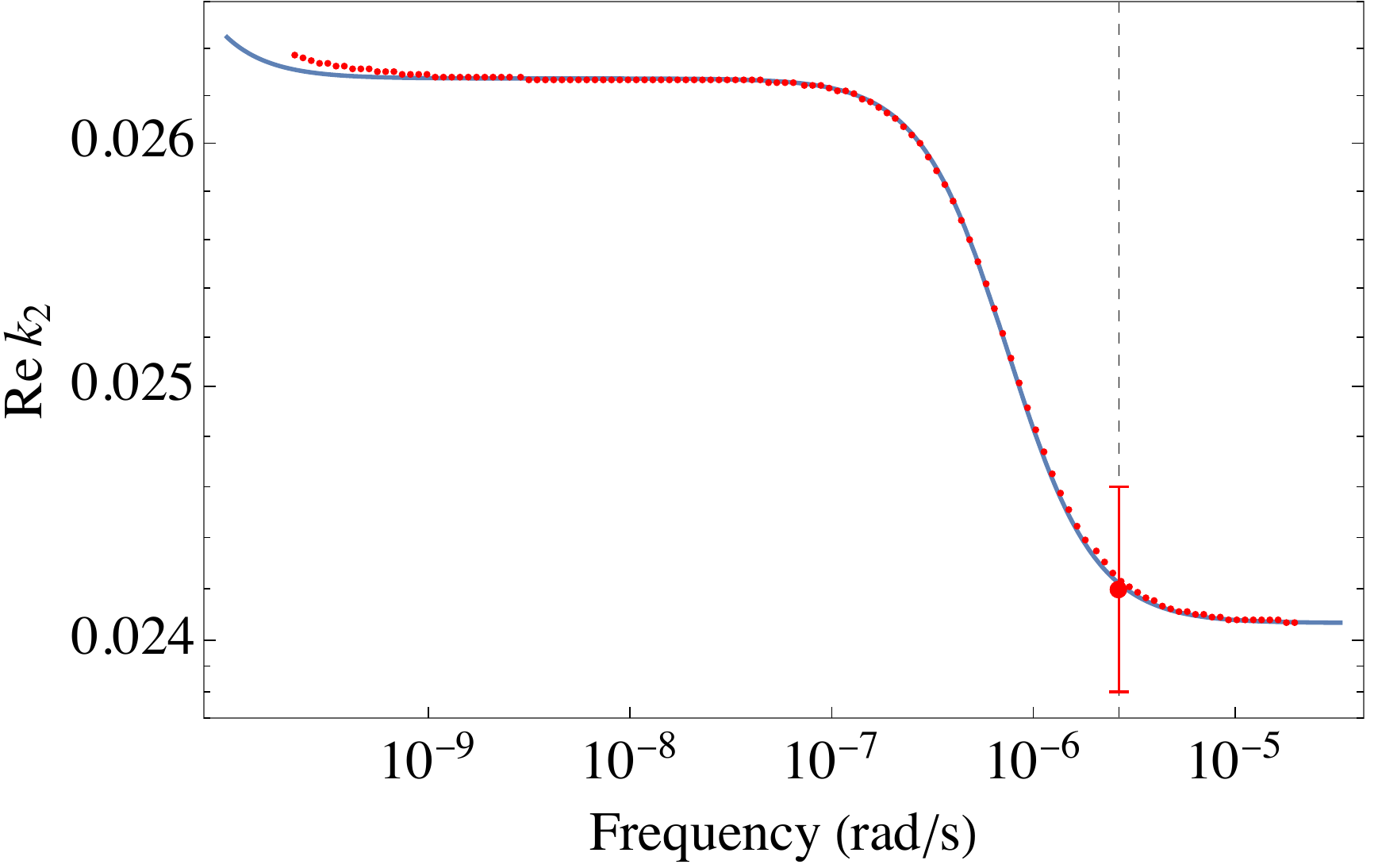}\\
\vspace{2em}
\includegraphics[scale=0.46]{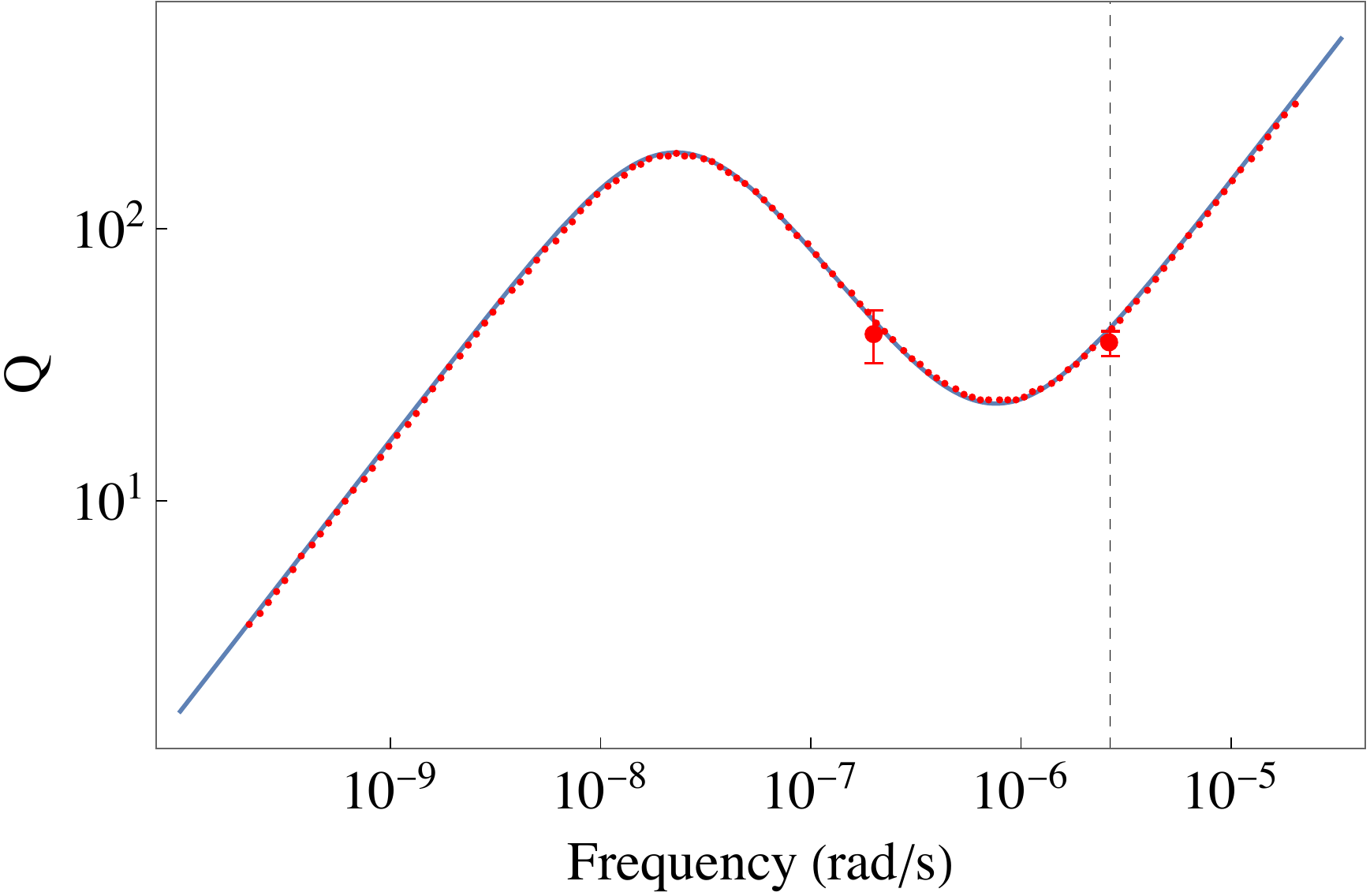}
\caption{Tidal Love number and quality factor dependence on frequency for the Moon. Red dotted lines show the imaginary and real parts of $k_2$ and the quality factor $Q$ as a function of frequency for the interior structure summarized in Table \ref{moon_int}. The red points with error bars are the observational constraints. Blue solid lines are obtained with homogeneous rheological models in Sections \ref{simplification} and \ref{manual}.\label{fig:fit}}
\end{figure}

 The tidal quality factor $Q$ is constrained at multiple forcing periods by lunar laser ranging \citep{Williams2015}: $Q=38\pm4$ at 1 month, $Q=41\pm9$ at 1 year, $Q\geq74$ at 3 years, and $Q\geq58$ at 6 years. Although the large uncertainties allow for a $Q$ decreasing (i.e. more dissipation) with forcing period, the mean values of the empirical $Q$ seems to increase (i.e. less dissipation) with forcing period. We adjust the low viscosity layer rigidity and viscosity to satisfy the tidal $Q$ constraints. This requires viscosities $\sim10^{16}$ Pa s, which is significantly smaller than typical Earth's mantle viscosities, as found in previous studies \citep{Harada2014,Matsumoto.2015}. Red dotted lines in Figure \ref{fig:fit} show the imaginary and real parts of $k_2$ and the quality factor $Q$ as a function of frequency for the interior structure summarized in Table \ref{moon_int}. The points with error bars in Figure \ref{fig:fit} are the observational constraints.

\subsection{Dissipation in a homogeneous body}\label{Homogeneous}

Suppose we are given a spherically symmetric body with interior  stratification. Each layer 
of the body is homogeneous, incompressible,
and with  Maxwell rheology.  We will show
the existence of  a hypothetical  homogeneous body with  the same 
response to tidal forcing as the multilayered body. 
More precisely we will show that the Love numbers $k_2$  of both the multilayer and the homogeneous bodies have the same dependence on the frequency of tidal forcing. 
The rheology of the homogeneous body has to be sufficiently complex and it can 
be either given by a  generalized Maxwell model or by a generalized Voigt model.

The degree-2 tidal Love number in the frequency domain (the domain of the Laplace transform $s\in\mathbb C$) is related to a complex compliance
$C(s)$ of the whole system defined as (see \cite{mathews2002modeling} paragraph [21])
\begin{equation}\label{k2cs}
k_2(s):= \left(\frac{3\mathrm{I}_{\circ} G}{ R^5}\right) C(s),
\end{equation}
where $\mathrm{I}_{\circ}$ is the mean moment of inertia (the dimension of $\frac{3\mathrm{I}_{\circ} G}{ R^5}$ is 1/time$^2$ and the dimension of $C(s)$ is time$^2$). A similar expression for $k_{2}(\omega)$ was obtained in \citet{Correia2018}. If the inertia of deformation is neglected, as assumed in this paper, then
\begin{equation}\label{eq:K2(omega)}
k_{2}(i\omega) = \frac{3G\rm{I}_{o}}{R^5}\frac{1}{\gamma+\hat{J}^{-1}(i\omega)},
\end{equation}
where $\hat{J}(i\omega)$ is the complex compliance of the viscoelastic element and $\gamma$ is a gravitational rigidity parameter related to the fluid Love number $k_f$ as \citep[Eq. 1.8]{ragazzo2020theory}
\begin{equation}
\gamma=\frac{3\mathrm{I}_{\circ}G}{R^5}\frac{1}{k_f}.
\end{equation}
Fluid Love number can be approximated by $k_f\approx \frac{3}{2}\left(\frac{R_I}{R}\right)^5$, where inertial radius is defined by $R_I=\sqrt{\frac{5}{2} \frac{I_\circ}{m}}$.


\subsubsection{The generalized Maxwell model}\label{maxwell}


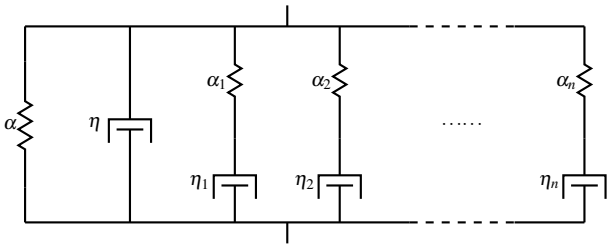
\begin{figure}
\begin{center}
\begin{tikzpicture}[scale=0.92, transform shape]
\tikzstyle{spring}=[thick, decorate, decoration={zigzag, pre length=0.5cm, post length=0.5cm, segment length=6}]
\tikzstyle{damper}=[thick, decoration={markings,
  mark connection node=dmp,
  mark=at position 0.5 with
  {
    \node (dmp) [thick, inner sep=0pt, transform shape, rotate=-90, minimum width=15pt, minimum height=3pt, draw=none] {};
    \draw [thick] ($(dmp.north east)+(5pt,0)$) -- (dmp.south east) -- (dmp.south west) -- ($(dmp.north west)+(5pt,0)$);
    \draw [thick] ($(dmp.north)+(0,-5pt)$) -- ($(dmp.north)+(0,5pt)$);
  }
}, decorate]
\tikzstyle{ground}=[fill,pattern=north east lines,draw=none,minimum width=0.75cm,minimum height=0.3cm]

            \draw [spring] (0,0.5) -- node[left] {$\alpha$} (0,2.3);
            \draw [thick] (0,0) -- (0,0.5);
            \draw [thick] (0,2.3) -- (0,2.8);
            \draw [thick] (0,2.8) -- (1.5,2.8);
            \draw [thick] (0,0) -- (1.5,0);            
            \draw [damper] (1.5,2.8) --  (1.5,0);
            \node at (1,1.4) {$\eta$};

           \draw [spring] (3,2.8) -- node[left] {$\alpha_1$} (3,1.2);
            \draw [damper] (3,1.2) -- (3,0);
            \node at (2.5,0.57) {$\eta_{1}$};
            
            \draw [thick] (3.75,2.8) -- (3.75,3.1);
            \draw [thick] (3.75,-0.3) -- (3.75,0);
           \draw [spring] (4.5,2.8) -- node[left] {$\alpha_2$} (4.5,1.2);
            \draw [damper] (4.5,1.2) -- (4.5,0);
            \node at (4,0.57) {$\eta_{2}$};

           \draw [spring] (8,2.8) -- node[left] {$\alpha_n$} (8,1.2);
            \draw [damper] (8,1.2) -- (8,0);
            \node at (7.5,0.57) {$\eta_{n}$};

            \draw [thick] (1.5,2.8) -- (5.5,2.8);
            \draw [thick] (1.5,0) -- (5.5,0);
            \draw [thick,dashed] (5.5,2.8) -- (7,2.8);
            \draw [thick,dashed] (5.5,0) -- (7,0);
            \draw [thick] (7,2.8) -- (8,2.8); \draw [thick] (7,0) -- (8,0);
            \node at (6.25,1.4) {$\ldots\!\ldots$};

      \end{tikzpicture}
\end{center}
\caption[General oscillator with rheology]{The generalized Maxwell model.}
\label{fig:max-osc}
\end{figure}

The following is true for the complex compliance of the generalized Maxwell model\footnote{The elastic parameters $\alpha,\alpha_1,\ldots\alpha_n$ have dimension of 1/time$^2$ and the viscosity parameters $\eta,\eta_1,\ldots\eta_n$ have dimension of 1/time. To go to the usual dimensions of $\rm{Pa}$ and $\rm{Pa}\cdot \rm{s}$ we should multiply them by the rescaling constant $\frac{15}{152\pi}\frac{m_{Moon}}{R}$ (for details see \citep{Correia2018})\label{footnote1}} in Figure \ref{fig:max-osc}: 

\begin{itemize}
\item[a)] The complex compliance of the model is given by \citep{bla2016}
\begin{equation}\label{C1}
C^{-1}(s)= \alpha+\eta s +\sum_{i=1}^n\left(\frac{1}{\alpha_i}+\frac{1}{\eta_i s}\right)^{-1}.
\end{equation}
\item[b)] Straightforward algebraic manipulations allow us to rewrite the complex compliance in the form
\begin{equation}\label{C2}
C(s):=\frac{P_2(s)}{P_1(s)}=\frac{(\eta_1\ldots\eta_n)(s+\omega_1)\ldots(s+\omega_n)}{P_1(s)},
\end{equation}
where $\omega_i^{-1}=\eta_i/\alpha_i$ is the Maxwell time of the $i$-th  Maxwell element, 
\begin{equation}\label{Ps}
P_1(s)=(\eta_1\ldots\eta_n)\left\{(\eta s +\alpha)\prod_{i=1}^n(\omega_i+s)+s\sum_{i=1}^n\alpha_i\prod_{j\ne i}^n(\omega_j+s)\right\},
\end{equation}
and where for later convenience we consider the $\omega_{i}$ ordered like $0<\omega_1 < \omega_2 < \cdots < \omega_n$.
\item[c)] $P_1(s)$ is a polynomial of degree $n+1$ with distinct real roots and can be written as
\begin{equation}\label{C22}
P_1(s)=\eta (\eta_1\ldots\eta_n)(s-s_1)\ldots(s-s_{n+1}),
\end{equation}
where $s_i$ is the inverse of the relaxation time of mode $i$.\footnote{The compliance $C(s)$ can be viewed as the transfer function of the homogeneous linear system composed by the elements of the generalized Maxwell rheology \cite[Eq.~5.31]{ragazzo2022librations}. The poles $s_i$ of $C(s)$ are, therefore, the roots of the characteristic equation, i.e., the eigenvalues of the linear system.}
From expression (\ref{Ps}) we see that $\sgn\{P_1(-\omega_i)P_1(-\omega_{i+1})\}=\sgn\{(-1)^{i}(-1)^{i+1}\}$ is negative, $P_1(0)=\alpha(\omega_1\omega_2\cdots\omega_n) > 0$ and $\sgn\{P_1(-\infty)\}=\sgn\{(-1)^{n+1}\}$ is opposite to $\sgn\{P_1(-\omega_n)\}$, meaning that $P_1(s)$ changes sign $n+1$ times along the negative real line. Since a continuous function that changes sign inside an interval has a root in that interval (Bolzano's theorem), $P_1(s)$ has a root in each of the intervals $(-\infty,-\omega_n)$, $(-\omega_{n},-\omega_{n-1})$, \dots, $(-\omega_1,0)$, totaling $n+1$ distinct negative real roots. The roots of $P_1$ and $P_2$ are related by (see Figure \ref{fig:roots_max})
\begin{equation}\label{order1}
0 <-s_1 < \omega_1 < -s_2 < \omega_2 < \cdots < -s_{n} < \omega_n < -s_{n+1}.
\end{equation}
\color{black}

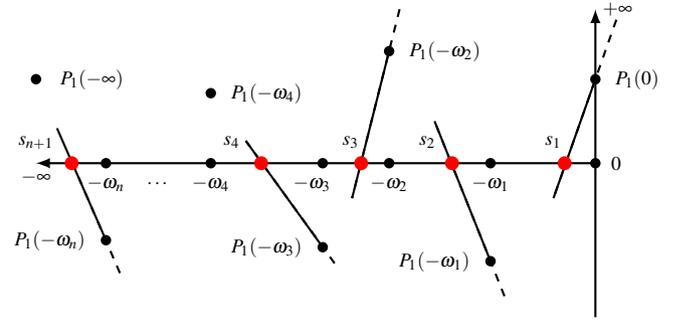
\begin{figure}
\begin{center}
\begin{tikzpicture}[scale=0.92, transform shape]
\tikzstyle{ground}=[fill,pattern=north east lines,draw=none,minimum width=0.75cm,minimum height=0.3cm]

\draw [-latex, thick] (8,0.7) -- (0,0.7)  node[below] {$-\infty$};
\draw [-latex, thick] (8,-1.5) -- (8,2.9)  node[right] {$+\infty$};

\node at (8,0.7) [circle,fill,inner sep=1.5pt]{};
\node at (8.3,0.7) {$0$};
\node at (8,1.9) [circle,fill,inner sep=1.5pt]{};
\node at (8.6,1.9) {$P_1(0)$};

\draw [thick, dashed] (8.3,2.75) -- (7.4,0.2);
\draw [thick] (8,1.9) -- (7.4,0.2);

\node at (7.56,0.7) [circle,fill,inner sep=2pt,red]{};
\node at (7.4,1) {$s_1$};

\node at (6.5,0.7) [circle,fill,inner sep=1.5pt]{};
\node at (6.5,0.4) {$-\omega_1$};

\node at (6.5,-0.7) [circle,fill,inner sep=1.5pt]{};
\node at (5.7,-0.7) {$P_1(-\omega_1)$};

\draw [thick, dashed] (6.7,-1.2) -- (5.7,1.3);
\draw [thick] (6.5,-0.7) -- (5.7,1.3);

\node at (5.95,0.7) [circle,fill,inner sep=2pt,red]{};
\node at (5.6,1) {$s_2$};

\node at (5.05,0.7) [circle,fill,inner sep=1.5pt]{};
\node at (5.05,0.4) {$-\omega_2$};

\node at (5.05,2.3) [circle,fill,inner sep=1.5pt]{};
\node at (5.85,2.3) {$P_1(-\omega_2)$};

\draw [thick, dashed] (5.2,2.9) -- (4.52,0.2);
\draw [thick] (5.05,2.3) -- (4.52,0.2);

\node at (4.65,0.7) [circle,fill,inner sep=2pt,red]{};
\node at (4.5,1) {$s_3$};

\node at (4.1,0.7) [circle,fill,inner sep=1.5pt]{};
\node at (3.95,0.4) {$-\omega_3$};

\node at (4.1,-0.5) [circle,fill,inner sep=1.5pt]{};
\node at (3.3,-0.5) {$P_1(-\omega_3)$};

\draw [thick, dashed] (4.27,-0.73) -- (3,1.02);
\draw [thick] (4.1,-0.5) -- (3,1.02);

\node at (3.22,0.7) [circle,fill,inner sep=2pt,red]{};
\node at (2.8,1) {$s_4$};

\node at (2.5,0.7) [circle,fill,inner sep=1.5pt]{};
\node at (2.5,0.4) {$-\omega_4$};

\node at (2.5,1.7) [circle,fill,inner sep=1.5pt]{};
\node at (3.3,1.7) {$P_1(-\omega_4)$};

\node at (1.75,0.4) {$\cdots$};

\node at (1,0.7) [circle,fill,inner sep=1.5pt]{};
\node at (1,0.4) {$-\omega_n$};

\node at (1,-0.4) [circle,fill,inner sep=1.5pt]{};
\node at (0.2,-0.4) {$P_1(-\omega_n)$};

\draw [thick, dashed] (1.2,-0.87) -- (0.5,0.75);
\draw [thick] (1,-0.4) -- (0.3,1.2);

\node at (0.51,0.7) [circle,fill,inner sep=2pt,red]{};
\node at (0,1) {$s_{n+1}$};

\node at (0,1.9) [circle,fill,inner sep=1.5pt]{};
\node at (0.8,1.9) {$P_1(-\infty)$};

\end{tikzpicture}
\end{center}
\caption[Root distribution]{Distribution of poles and zeros of the complex compliance. The figure depicts the case for $n$ odd.}
\label{fig:roots_max}
\end{figure}
\item[d)] The decomposition in partial fractions of $C(s)$ is given by \begin{equation}\label{part1}
C(s)=\frac{1}{\eta}\left(\frac{A_1}{s-s_1}+\cdots+\frac{A_{n+1}}{s-s_{n+1}}\right),
\end{equation}
where 
\begin{equation}
A_i = \frac{1}{1/\eta}\times\frac{P_2(s_i)}{P_1'(s_i)}=\frac{(s_i+\omega_1)\ldots(s_i+\omega_n)}{\prod\limits_{j\ne i}^ {n+1}(s_i-s_j)} > 0,
\end{equation}
where the positivity of $A_i$ comes from (\ref{order1}), which gives $\sgn\{P_2(s_i)\} = \sgn\{(-1)^{i-1}\} = \sgn\{P_1'(s_i)\}$, and $\sum_{i}A_i=1$, since from equations (\ref{C2}) and (\ref{C22}) we have
$\lim_{s\to\infty}\,sP_2(s)/P_1(s) = 1/\eta$ and from equation (\ref{part1}) we have $\lim_{s\to\infty}\,sC(s) = (1/\eta)\sum_{i}A_i$, furnishing the result.
\end{itemize}
\color{black}
The second degree tidal Love number can be rewritten using equations (\ref{k2cs}) and (\ref{part1}), as
\begin{equation}\label{k2maxwell}
k_2(s)=\left(\frac{3\mathrm{I}_{\circ} G}{ R^5}\right)\times\frac{1}{\eta}\left(\frac{A_1}{s-s_1}+\cdots+\frac{A_{n+1}}{s-s_{n+1}}\right).
\end{equation}

\subsubsection{The generalized Voigt model}\label{gen_voi}


\begin{figure}
\begin{center}
\begin{tikzpicture}[scale=0.92, transform shape]
\tikzstyle{spring}=[thick, decorate, decoration={zigzag, pre length=0.5cm, post length=0.5cm, segment length=6}]
\tikzstyle{damper}=[thick, decoration={markings,
  mark connection node=dmp,
  mark=at position 0.5 with
  {
    \node (dmp) [thick, inner sep=0pt, transform shape, rotate=-90, minimum width=15pt, minimum height=3pt, draw=none] {};
    \draw [thick] ($(dmp.north east)+(5pt,0)$) -- (dmp.south east) -- (dmp.south west) -- ($(dmp.north west)+(5pt,0)$);
    \draw [thick] ($(dmp.north)+(0,-5pt)$) -- ($(dmp.north)+(0,5pt)$);
  }
}, decorate]
\tikzstyle{ground}=[fill,pattern=north east lines,draw=none,minimum width=0.75cm,minimum height=0.3cm]

            \draw [thick] (0,0.8) --  (0,2.5);

           \draw [spring] (0,0.8) -- node[below] {$\alpha$} (2,0.8);
            \draw [damper] (1.5,0.8) -- (3,0.8);
            \node at (2.4,0.3) {$\eta$};

            \draw [thick] (3,0.2) --  (3,1.4);

           \draw [spring] (3,1.4) -- node[above] {$\alpha_1$} (4.3,1.4);
            \draw [damper] (3,0.2) -- (4.3,0.2);
            \node at (3.75,-0.3) {$\eta_{1}$};

            \draw [thick] (4.3,0.2) --  (4.3,1.4);

            \draw [thick] (4.3,0.8) --  (4.8,0.8);
            \draw [thick,dashed] (4.8,0.8) -- (5.7,0.8);
            \draw [thick] (5.7,0.8) -- (6.2,0.8);

            \draw [thick] (6.2,0.2) --  (6.2,1.4);
           \draw [spring] (6.2,1.4) -- node[above] {$\alpha_n$} (7.5,1.4);
            \draw [damper] (6.2,0.2) -- (7.5,0.2);
            \node at (6.95,-0.3) {$\eta_{n}$};

            \draw [thick] (7.5,0.2) --  (7.5,1.4);

            \draw [thick] (7.5,0.8) --  (8,0.8);

            \draw [thick] (8,0.8) --  (8,2.5);

            \draw [thick] (0,2.5) --  (2.5,2.5);
           \draw [spring] (2.5,2.5) -- node[above] {$\gamma$} (5.5,2.5);
            \draw [thick] (5.5,2.5) --  (8,2.5);

            \draw [thick] (-0.3,1.65) --  (0,1.65);
            \draw [thick] (8,1.65) --  (8.3,1.65);

      \end{tikzpicture}
\end{center}
\caption[General oscillator with rheology]{The generalized Voigt model.}
\label{fig:voig-osc}
\end{figure}
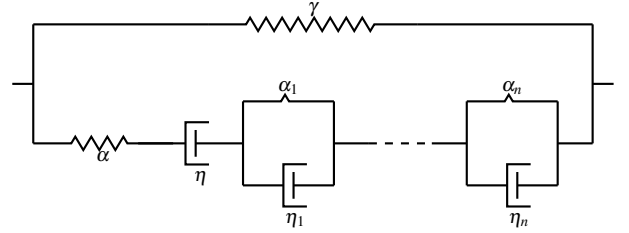
The following is true for the complex compliance of the generalized Voigt model with the additional spring $\gamma$ in  parallel as in Figure \ref{fig:voig-osc}:

\begin{itemize}
\item[a)] The complex compliance of the model is given by \citep{bla2016}
\begin{equation}\label{CV0}
C^{-1}(s)=\gamma +\bigg\{
\frac{1}{\alpha}+\frac{1}{\eta s}+\sum_{i=1}^n\frac{1}{\alpha_i+s\eta_i}
\bigg\}^{-1}
\end{equation}
and 
\begin{equation}\label{CV1}
C(s):= \frac{1}{\alpha+\gamma}+ C_v(s),
\end{equation}
where $C_v(s)$ is the viscous compliance as defined in \cite[equation (1.192)]{sabadini2016global}. The following limits hold:
\begin{equation}
\lim_{s\to\infty}C(s)=\frac{1}{\gamma+\alpha}\quad \text{and}\quad
\lim_{s\to\infty}s C_v(s)=\frac{\alpha^2}{(\gamma+\alpha)^2}
\left(\frac{1}{\eta}+\sum_{i=1}^n\frac{1}{\eta_i}\right)\label{Vlim}
\end{equation}
\item[b)] Straightforward algebraic manipulations allow us to rewrite the viscous compliance in the form
\begin{equation}
C_v(s):=\frac{P_2(s)}{P_1(s)},
\end{equation}
where 
\begin{equation}
P_2(s) = \alpha^{2}(\eta_1\ldots\eta_n)\left\{(\prod_{i=1}^n(\omega_i+s)+\eta s\sum_{i=1}^n\frac{1}{\eta_i}\prod_{j\ne i}^n(\omega_j+s)\right\},
\end{equation}

\begin{equation}
\begin{split}
&P_1(s)=(\alpha+\gamma)(\eta_1\ldots\eta_n)\\
&\times\left\{(\alpha\eta s + \gamma\eta s +\gamma\alpha)\prod_{i=1}^n(\omega_i+s)+\gamma\alpha\eta s\sum_{i=1}^n\frac{1}{\eta_i}\prod_{j\ne i}^n(\omega_j+s)\right\}\\
&=(\alpha+\gamma)(\eta_1\ldots\eta_n)\left\{\eta(\alpha + \gamma)s\prod_{i=1}^n(\omega_i+s)+\frac{\gamma}{\alpha}P_{2}(s)\right\},
\end{split}
\end{equation}
$\omega_i^{-1}=\eta_i/\alpha_i$ is the characteristic time of $i-$th Kelvin-Voigt element, and for later convenience we consider the $\omega_{i}$ ordered like $0<\omega_1 < \omega_2 < \cdots < \omega_n$.

\item[c)] $P_{2}(s)$ is a polynomial of degree $n$ with distinct negative real roots and can be written as
\begin{equation}
P_2(s)=\alpha^{2}(\eta_1\ldots\eta_n)(1+\eta\sum_{i=1}^n\frac{1}{\eta_i})(s-q_1)\ldots(s-q_{n}).
\end{equation}
The roots of $P_2(s)$ are related to the inverse characteristic times of Kelvin-Voigt elements as
\begin{equation}\label{Qin}
0<-q_1<\omega_1<-q_2<\omega_2<\cdots<-q_n<\omega_n.
\end{equation}
$P_1(s)$ is a polynomial of degree $n+1$ with distinct real roots and can be written as 
\begin{equation}
P_1(s)=(\alpha+\gamma)^{2}\eta (\eta_1\ldots\eta_n)(s-s_1)\ldots(s-s_{n+1}).
\end{equation}
The roots of $P_1(s)$ are related to the inverse characteristic times of Kelvin-Voigt elements as
\begin{equation}
0 <-s_{1} < -q_{1}< -s_{2} < -q_{2} < \cdots < -s_{n} < -q_{n} < -s_{n+1}.
\end{equation}
\item[d)] The decomposition in partial fractions of $C_v(s)$ is given by
\begin{equation}\label{part2}
C_v(s)=\frac{\alpha^2}{(\gamma+\alpha)^2}
\left(\frac{1}{\eta}+\sum_{i=1}^n\frac{1}{\eta_i}\right)
\left(\frac{A_1}{s-s_1}+\cdots+\frac{A_{n+1}}{s-s_{n+1}}\right),
\end{equation}
where
\begin{equation}
A_i = \frac{1}{\dfrac{\alpha^2}{(\gamma+\alpha)^2}
\left(\dfrac{1}{\eta}+\sum\limits_{i=1}^n\dfrac{1}{\eta_i}\right)}\times\frac{P_2(s_i)}{P_1'(s_i)}=\frac{(s_i-q_1)\ldots(s_i-q_n)}{\prod\limits_{j\ne i}^ {n+1}(s_i-s_j)} > 0
\end{equation}
and $\sum_{i=1}^{n+1}A_i=1$. The facts presented above can be obtained if we proceed in a manner similar to that in Subsection \ref{maxwell} for the generalized Maxwell model.
\end{itemize}
\color{black}
The second degree tidal Love number can be rewritten using equations (\ref{k2cs}), (\ref{CV1}) and (\ref{part2}), as
\begin{equation}\label{k2voigt}
\begin{split}
&k_2(s) = \left(\frac{3\mathrm{I}_{\circ} G}{ R^5}\right)\\
&\times\Bigg(\frac{1}{\alpha+\gamma} + \frac{\alpha^2}{(\gamma+\alpha)^2}
\left(\frac{1}{\eta} + \sum_{i=1}^n\frac{1}{\eta_i}\right)
\left(\frac{A_1}{s-s_1} + \cdots+\frac{A_{n+1}}{s-s_{n+1}}\right)\Bigg).
\end{split}
\end{equation}
The only dynamical difference between generalized Voigt and generalized Maxwell models is the asymptotic behavior at high frequencies,  i.e. as $|s|\to\infty$. At high frequencies, a body with a generalized Voigt rheology  behaves as a purely elastic body while one with a generalized Maxwell rheology behaves as a rigid body. 

Note that in the limit $n\to\infty$, a generalized Voigt rheology is equivalent to an Andrade or a Sundberg-Cooper rheology, frequently used to model otherwise stratified moons and planets \citep{gev2020, Gev2021}.

    
\subsection{Normal modes of stratified incompressible rheological models}\label{normal_modes}

For a stratified body treated as in Section \ref{stratified} the second degree tidal Love number $k_2(s) = N(s)/D(s)$ is a ratio of two polynomials. The roots of the secular equation $D(s)=0$ and the Love number $k_2(s)$ have the following properties (most of them can be seen in \citep[p. 38]{sabadini2016global} and \citep[the paragraph between equations (55) and (56)]{wu1982viscous}):
\begin{itemize}
\item[a)] All  roots of the secular equation are on the real axis \citep{tanaka2006new}.
\item[b)] The secular equation has a finite number of roots, since $D(s)$ is a polynomial. 
\item[c)] All roots of the secular equation are simple and negative: $0 > s_1 > s_2 >\ldots > s_{n}$, where $n$ is the degree of D, unless for exceptional layered structures, e.g. when the density of a layer is lower than that of the neighboring layer above. The negativity of the roots is a necessary condition for the stability with respect to tidal forcing.
\item[d)] The tidal Love number $k_2(s)$  can be  decomposed into  partial fractions, as in \citep[pg. 105 equation (3.67)]{sabadini2016global}: 
\begin{equation}\label{saba367}
k_2(s)=k_E+\sum_{i=1}^n\frac{r_i}{s-s_i}=k_E+\sum_{i=1}^n\tau_ir_i\frac{1}{\tau_i s +1},
\end{equation}
where $k_E$ is the elastic tidal Love number that characterizes the behavior of the body as $s\to\infty$, $r_i$ 
is the amplitude, and 
$\tau_i=-1/s_i$ is the relaxation time of each mode $i$.
\item[e)] The amplitudes $r_i$ of the modes are positive. Let $s=i \omega$, where $\omega>0$ is the tidal forcing frequency. Each term in the partial fraction expansion (\ref{saba367}) can be decomposed into real and complex parts as
\begin{equation}
r_j\frac{1/\tau_j-i\omega}{\omega^2+1/\tau_j^2}.
\end{equation}
Since  $\tau_j>0$ and $\omega>0$,  the real part is positive (inertial effects were neglected) and the imaginary part is negative (energy is dissipated) if, and only if,  the amplitude $r_j$ is positive.
\end{itemize}

We will say that a body has a {\it simple layered rheology} if it has finite characteristic times and if its Love number has the properties listed above, which are natural and plausible. A  body with finite homogeneous layers, each one with Maxwell rheology, is an example of a body with simple layered rheology. Note that the Maxwell rheology of each homogeneous layer can be replaced by other commonly used rheologies, e.g. Kelvin-Voigt, Burgers, or Andrade. If the number of normal modes is infinite, like for the Andrade model, we would need an infinite number of elements in the generalized Voigt rheology or, alternatively, we could use an effective Andrade or Sundberg-Cooper rheology as in \citep{Gev2021}.

Since equations (\ref{saba367}) and (\ref{k2maxwell}) have the same  structure if $k_E=0$ and  equations (\ref{saba367}) and (\ref{k2voigt}) have the same structure if $k_E>0$, the following result holds:

{ \it
The Love number $k_2(s)$ of a  body with simple layered rheology is equal to the Love number $k_2(s)$ of a hypothetical homogeneous body with generalized Maxwell or generalized Voigt rheology.
}

To any given simple layered rheology we can associate either a homogeneous generalized Maxwell or generalized Voigt rheology. The choice of one or another is a matter of convenience since the conditions $k_E=0$ (generalized Maxwell) or $k_E>0$ generalized Voigt can be easily changed either by making $\eta=0$ in the generalized Maxwell rheology, see equation (\ref{C1}) and Figure \ref{fig:max-osc}, or by making $\alpha=\infty$ in the generalized Voigt rheology, see equation (\ref{CV1}) and Figure \ref{fig:voig-osc}.

The formulas to go from $A_i$ to $r_i$ are: for the generalized Maxwell model,
\begin{equation}
r_i=\left(\frac{3\mathrm{I}_{\circ} G}{ R^5}\right)\frac{1}{\eta}\times A_i,
\end{equation}
and for the generalized Voigt model
\begin{equation}\label{fit_voigt}
\begin{split}
r_i&=\left(\frac{3\mathrm{I}_{\circ} G}{ R^5}\right)\frac{\alpha^2}{(\gamma+\alpha)^2}\left(\frac{1}{\eta}+\sum_{i=1}^n\frac{1}{\eta_i}\right)\times A_i, \\ k_E&=\left(\frac{3\mathrm{I}_{\circ} G}{R^5}\right)\frac{1}{\gamma+\alpha}.
\end{split}
\end{equation}

\subsection{Normal mode solution for the stratified Moon} 

Here we present the normal mode decomposition for the stratification we chose for the Moon in  Section \ref{stratified}. Real and imaginary parts of the second degree tidal Love number are
\begin{equation}\label{realimag}
\Re[k_2]=k_E-\sum_{i=1}^{N}\frac{s_ir_i}{s_i^2+\omega^2},\qquad \Im[k_2]=-\sum_{i=1}^{N}\frac{\omega r_i}{s_i^2+\omega^2}.
\end{equation}
The normal mode amplitudes, inverse relaxation times and elastic tidal Love number are given in Table~\ref{k2_expansion}. 
\begin{table*}
\begin{center}
\caption{The inverse relaxation time, actual and normalized amplitude of each mode $i$.}
\begin{tabular}{c c c c}
\hline
i \hspace{1.0cm} & $s_i (\rm{s}^{-1})$ \hspace{1.0cm} & $r_i$ & $\tau_ir_i$ \\
\hline
1 \hspace{1.0cm} & $-0.1$ \hspace{1.0cm} & $2.3156243820\times 10^{-17}$ & $2.3156243820\times 10^{-16}$\\
2 \hspace{1.0cm} & $-0.1$ \hspace{1.0cm} & $9.8181642643\times 10^{-17}$ & $9.8181642643\times 10^{-16}$\\
3 \hspace{1.0cm} & $-0.1$ \hspace{1.0cm} & $5.2900155892\times 10^{-17}$ & $5.2900155892\times 10^{-16}$\\
4 \hspace{1.0cm} & $-0.1$ \hspace{1.0cm} & $8.9387037631\times 10^{-16}$ & $8.9387037631\times 10^{-15}$\\
\textcolor{red}{5} \hspace{1.0cm} & $\textcolor{red}{-9.4591261374\times 10^{-7}}$ \hspace{1.0cm} & $\textcolor{red}{8.1074541438\times 10^{-10}}$ & $\textcolor{red}{0.86\times10^{-3}}$\\
\textcolor{red}{6} \hspace{1.0cm} & $\textcolor{red}{-6.3419112927\times 10^{-7}}$ \hspace{1.0cm} & $\textcolor{red}{8.5289052773\times 10^{-10}}$ &$\textcolor{red}{1.35\times 10^{-3}}$\\
7 \hspace{1.0cm} & $-5.0230520036\times 10^{-10}$ \hspace{1.0cm} & $2.8740137719\times 10^{-14}$ &$0.57\times 10^{-4}$\\
8 \hspace{1.0cm} & $-1.7092727755\times 10^{-12}$ \hspace{1.0cm} & $6.2186426095\times 10^{-15}$ &$3.64\times 10^{-3}$\\
\textcolor{red}{9} \hspace{1.0cm} & $\textcolor{red}{-1.3953038400\times 10^{-12}}$ \hspace{1.0cm} & $\textcolor{red}{1.5631724063\times 10^{-12}}$ &$\textcolor{red}{1.05}$\\
10 \hspace{1.0cm} & $-1.2295875216\times 10^{-13}$ \hspace{1.0cm} & $3.4739850530\times 10^{-14}$ &$0.283$\\
11 \hspace{1.0cm} & $-2.4641040361\times 10^{-14}$ \hspace{1.0cm} & $8.1522986569\times 10^{-17}$ &$3.31\times 10^{-3}$\\
12 \hspace{1.0cm} & $-2.4038824848\times 10^{-14}$ \hspace{1.0cm} & $2.1438046989\times 10^{-17}$ &$0.89\times 10^{-3}$\\
13 \hspace{1.0cm} & $-7.9022589544\times 10^{-17}$ \hspace{1.0cm} & $4.0680469600\times 10^{-20}$ &$0.52\times 10^{-3}$\\
\hline
\end{tabular} \\[0.3em]
{\footnotesize 
The elastic tidal Love number is $k_E=0.024068$. The terms in red are dominant and are the only ones needed to reproduce the tidal Love number frequency dependence seen in Figure \ref{fig:k2_exp}}
\label{k2_expansion}
\end{center}
\end{table*}
The number of modes depends on the stratification choice for the body \citep{sabadini2016global} and can be relatively large. Large the number of modes requires large number of building elements for the homogeneous rheology, hence we end up with a complex problem to fit the parameters. In the next section we present a way to reduce the number of parameters.


\section{Simplified homogeneous rheological model}\label{Simplified}

Here we present and compare two procedures to reduce the number of parameters and consequently to simplify the homogeneous rheology: either we find and consider only the dominant modes, which reduces the number of parameters, hence the number of elements in the generalized Maxwell or Voigt model, or we simply fit the Love number frequency dependence curve by hand using a simple rheology with few parameters. The second approach is always feasible, the first one can not be applied if there is not enough information on the internal structure of the body or if the number of dominant modes is too large.  

\subsection{Simplification by finding dominant modes}\label{simplification}

The fact that $\sum_{i=1}^n A_i=1$ and $A_i>0$, which holds for both the generalized Maxwell and Voigt rheologies, indicates that some of the modes in Sections \ref{maxwell} and \ref{gen_voi} must be dominant. The same is true for bodies with  stratification  even when they have infinitely many characteristic times. Indeed, in the limit as $n\to\infty$ this would imply that $A_n\to 0$, and so the smallness of $A_i$ in physically reasonable situations must hold  for $n$ large but finite. 
We show in Appendix \ref{modes} that effective rheologies may be constructed with just dominant amplitudes. Following the procedure in the appendix we conclude that terms marked in red in Table \ref{k2_expansion} are dominant in the frequency interval from $2\times10^{-5}$ to $2\times10^{-10}$ rad/s, hence we can omit rest of the terms, with no change in tidal Love number frequency dependence, the result is illustrated in Figure \ref{fig:k2_exp}. The number of relevant modes in the chosen frequency interval is reduced to three. Since we have three dominant modes, we need only two Voigt elements in the generalized Voigt model. We can now fit the parameters of generalized Voigt rheology from the data points 5, 6 and 9 in Table \ref{k2_expansion}, and the equations (\ref{fit_voigt}). We obtain the following values for the parameters of generalized Voigt rheology (we will call this rheology 1st rheology): $\gamma=1.02159\;\mathrm{GPa}$, $\alpha=60\;\mathrm{GPa}$, $\eta=9.05\times 10^{20}\;\mathrm{Pa\cdot s}$, $\alpha_1= 1050\;\mathrm{GPa}$, $\eta_1=1.66\times 10^{18} \;\mathrm{Pa\cdot s}$, $\alpha_2= 1662\;\mathrm{GPa}$ and $\eta_2=1.76\times 10^{18} \;\mathrm{Pa\cdot s}$. The fit obtained with these parameters can be seen in Figure \ref{fig:fit}.

The discussion above shows that a simplification of a rheological model depends on the choice of celestial object and the timescale one is interested in. In this paper we chose the Moon, the second best studied object in the Solar System, to present a procedure to simplify the homogeneous rheology associated to a stratified body. The same procedure  can be applied to other bodies.


\subsection{Manual simplification}\label{manual}

\begin{figure}
\begin{center}
\begin{tikzpicture}[scale=0.92, transform shape]
\tikzstyle{spring}=[thick,decorate,decoration={zigzag,pre length=0.5cm,post length=0.5cm,segment length=6}]
\tikzstyle{damper}=[thick,decoration={markings,  
  mark connection node=dmp,
  mark=at position 0.5 with 
  {
    \node (dmp) [thick,inner sep=0pt,transform shape,rotate=-90,minimum width=15pt,minimum height=3pt,draw=none] {};
    \draw [thick] ($(dmp.north east)+(5pt,0)$) -- (dmp.south east) -- (dmp.south west) -- ($(dmp.north west)+(5pt,0)$);
    \draw [thick] ($(dmp.north)+(0,-5pt)$) -- ($(dmp.north)+(0,5pt)$);
  }
}, decorate]
\tikzstyle{ground}=[fill,pattern=north east lines,draw=none,minimum width=0.75cm,minimum height=0.3cm]

            \draw [thick] (0,0.7) -- (1,0.7);
            \draw [thick] (1,0) -- (1,1.4);
            \draw [-latex, thick] (3,1.4) -- (0.5,1.4) node[above] {$f-s$};
            \draw [-latex, thick] (1,0) -- (0.5,0) node[below] {$s$};
            \draw [spring] (1,0) -- node[above] {$\alpha$} (3,0);
            \draw [damper] (3,0) -- (4,0);
            \node at (3.6,0.5) {$\eta$};
            \draw [thick] (4,0) -- (4.5,0);
            \draw [thick] (4.5,-0.5) -- (4.5,0.5);
            \draw [spring] (4.5,0.5) -- node[above] {$\alpha_{1}$} (6.5,0.5);
            \draw [damper] (4.5,-0.5) -- (6.5,-0.5);
            \node at (5.6,0) {$\eta_{1}$};
            \draw [thick] (6.5,-0.5) -- (6.5,0.5);            
            \draw [-latex,thick] (6.5,0) -- (8,0) node[below] {$s$};
            \draw [spring] (3,1.4) -- node[above] {$\gamma$} (5.5,1.4);
            \draw [-latex, thick] (5.5,1.4) -- (8,1.4) node[above] {$f-s$};
            \draw [thick] (7.5,0) -- (7.5,1.4);
            \draw [-latex, thick] (7.5,0.7) -- (9,0.7) node[below] {$f$};
            \draw [latex-latex, thick] (1,-1) -- node[below] {$x$} (7.5,-1);

\node (wall) at (-0.15,0.5) [ground, rotate=-90, minimum width=3cm] {};
\draw [thick] (wall.north east) -- (wall.north west);
      \end{tikzpicture}
\end{center}
\caption{Burgers oscillator. The external force $f(t)$ splits into the force $s(t)$ that acts upon Burgers array plus the force $f(t)-s(t)$ that acts upon the $\gamma$ spring.}
\label{burgers-osc}
\end{figure}
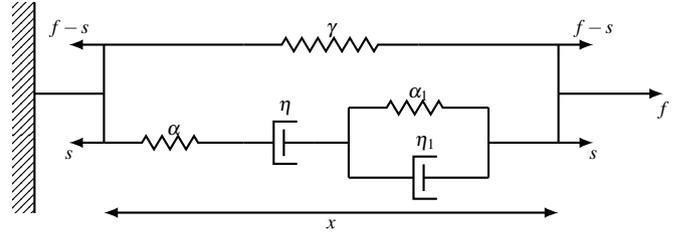

In this section we present a discussion on how to approximate tidal Love number frequency dependence given the plots in Figure \ref{fig:fit} or alike for any body. The simplest rheological model able to mimic the dissipation curve in the first plot of Figure \ref{fig:fit} is the well-known Burgers rheology in Fig. \ref{burgers-osc} \citep{Shoji2013}, which we will use to approximate the multilayered Moon by a homogeneous body.

\begin{figure*}
\centering
\begin{tabular}{cc}
a) $\alpha=60\cdot10^9$ Pa, $\eta_1=8.4\cdot10^{17}$ Pa s, $\eta=9\cdot10^{20}$ Pa s & b) $\alpha_1=61\cdot10^{10}$ Pa, $\eta_1=8.4\cdot10^{17}$ Pa s, $\eta=9\cdot10^{20}$ Pa s\\
\includegraphics[scale=0.46]{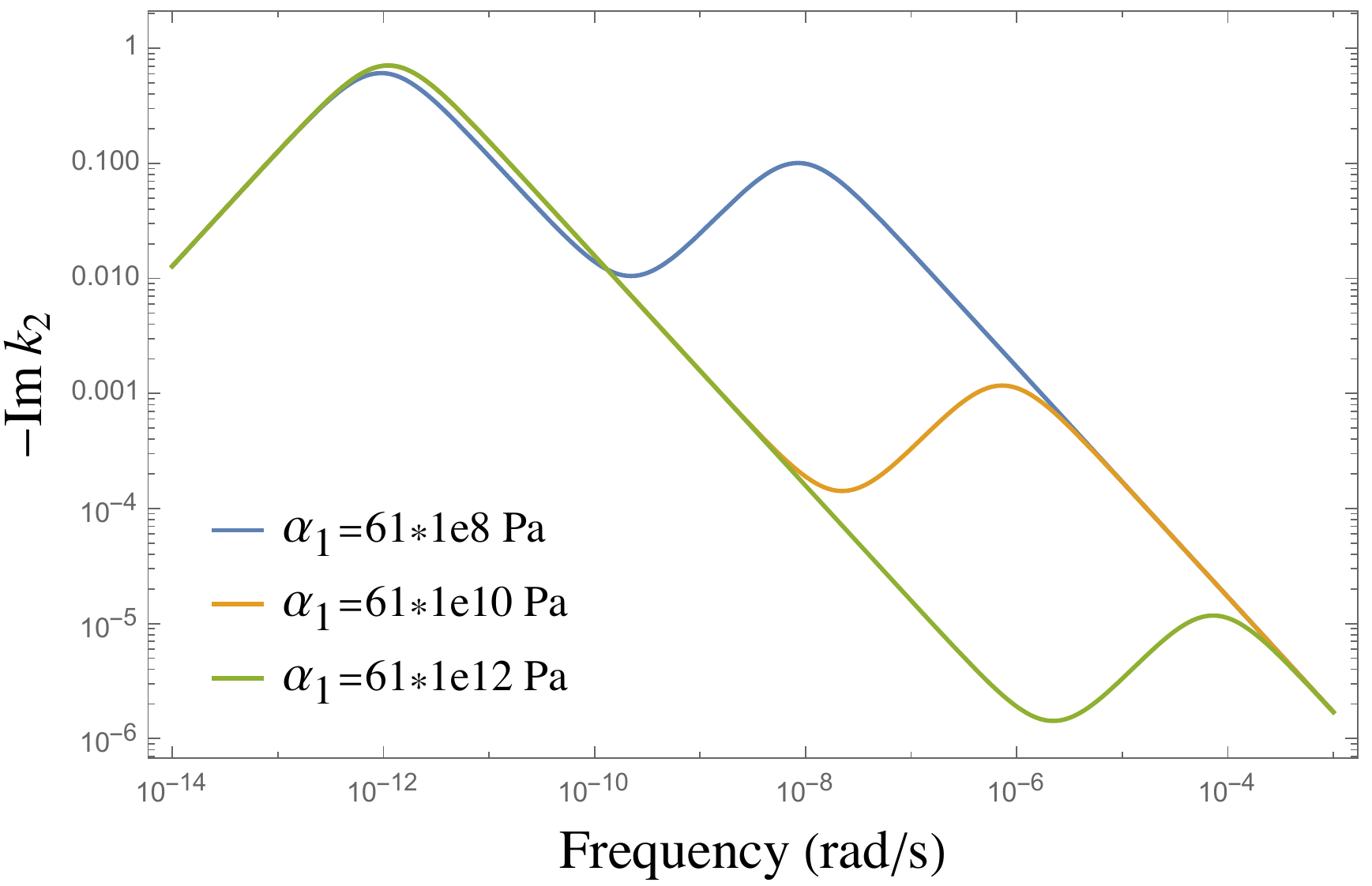} & \includegraphics[scale=0.46]{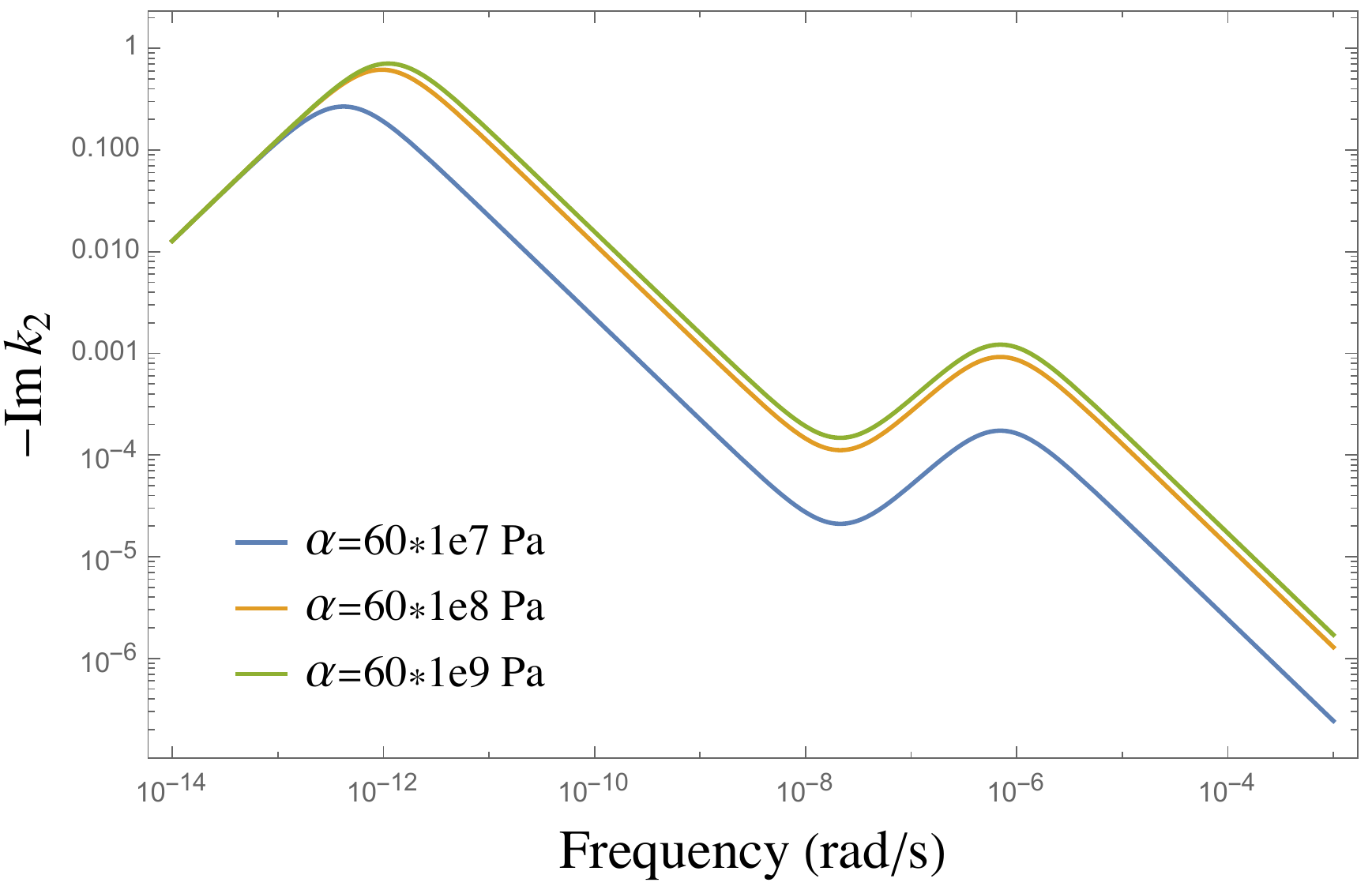}
\vspace{2em}\\
c) $\alpha_1=61\cdot10^{10}$ Pa, $\alpha=60\cdot10^9$ Pa, $\eta=9\cdot10^{20}$ Pa s & d) $\alpha_1=61\cdot10^{10}$ Pa, $\alpha=62\cdot10^9$ Pa, $\eta_1=8.4\cdot10^{17}$ Pa s\\
\includegraphics[scale=0.46]{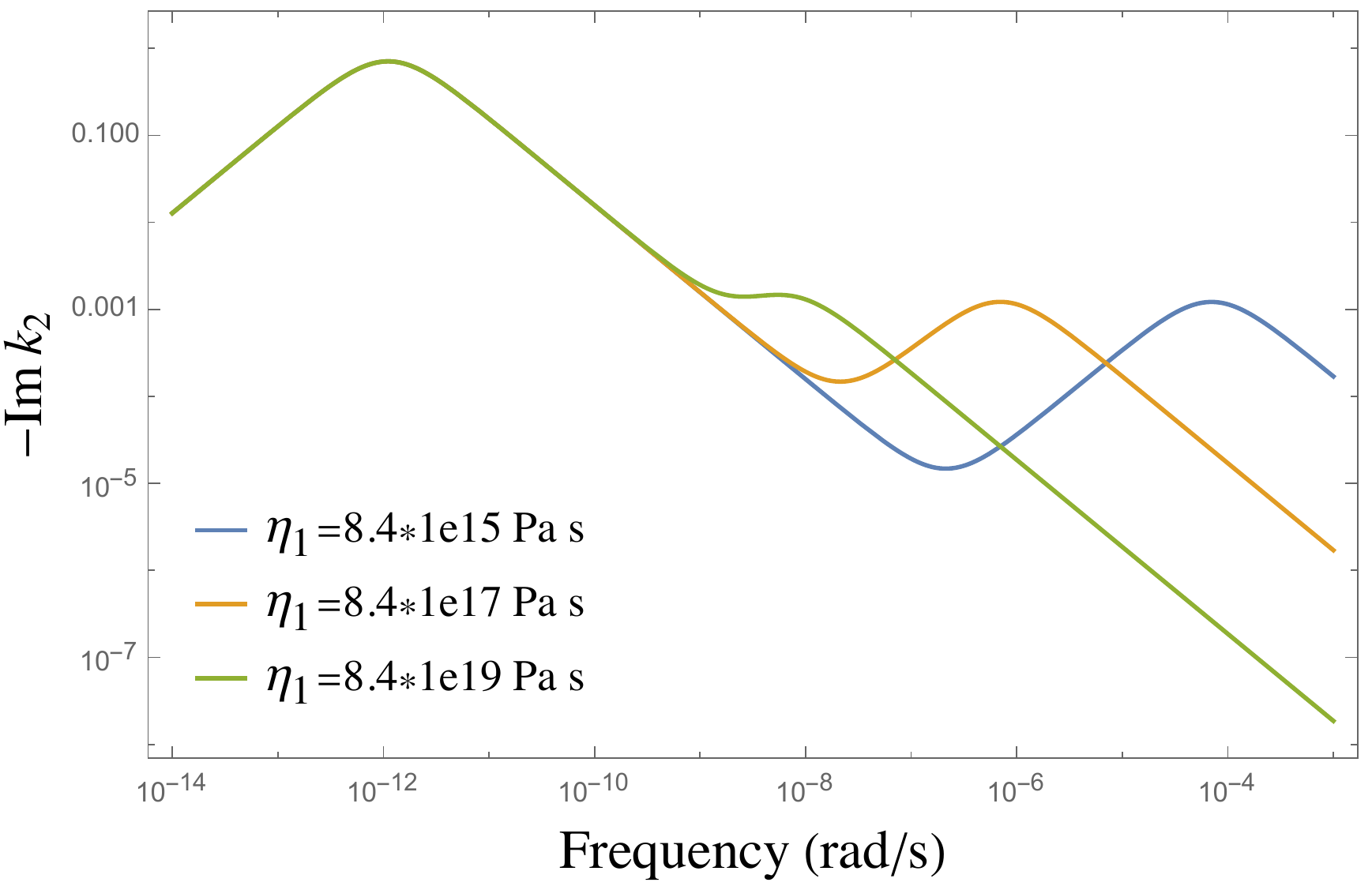} & \includegraphics[scale=0.46]{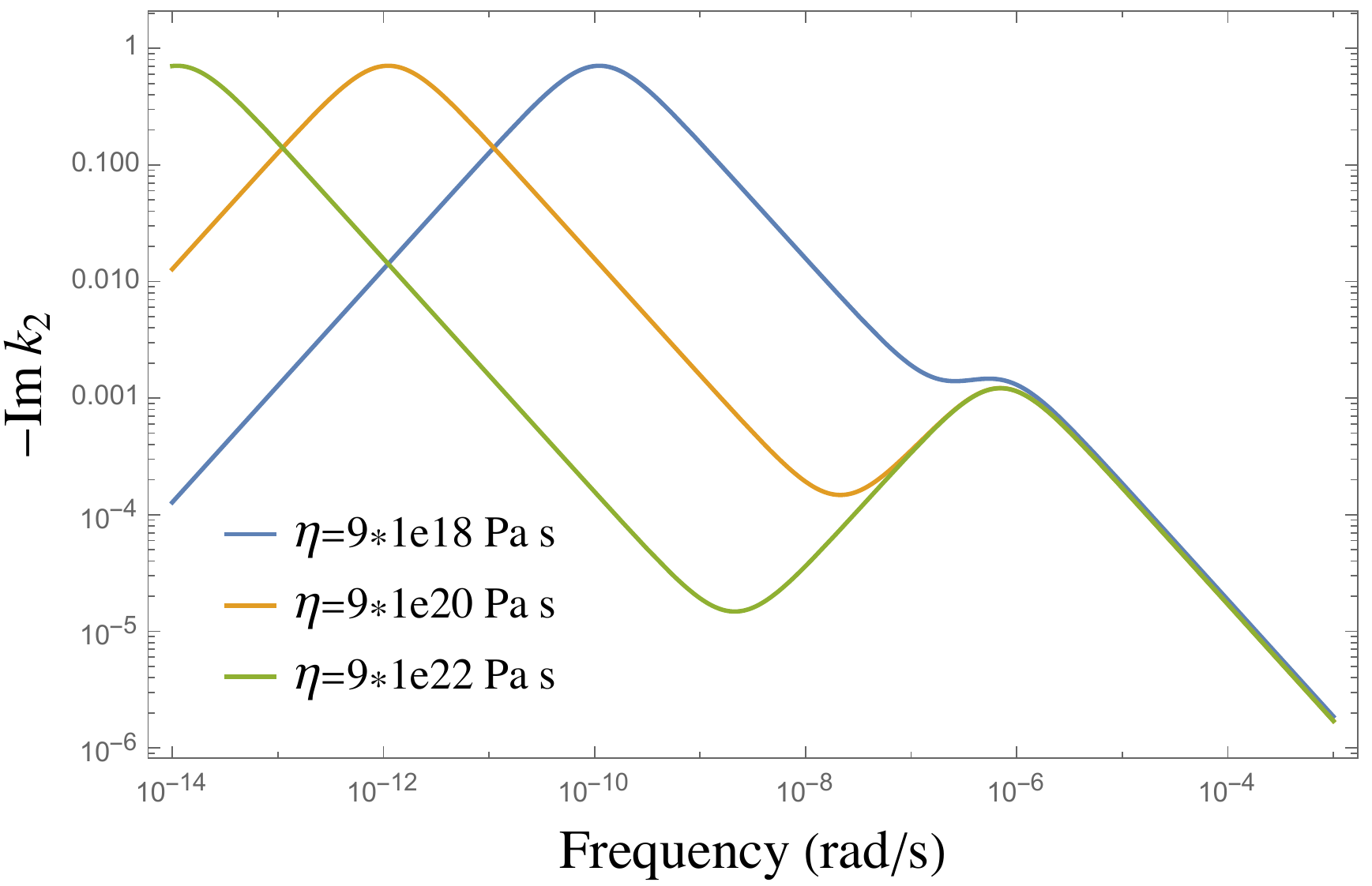}\\
\end{tabular}
\caption{Imaginary part of the degree-2 tidal Love number, which describes dissipation, as a function of frequency for a homogeneous body modeled with Burgers rheology; (a) effect of the Voigt element elasticity $\alpha_1$; (b) effect of the Maxwell element elasticity $\alpha$; (c) effect of the Voigt element viscosity $\eta_1$; (d) effect of the Maxwell element viscosity $\eta$.\label{fig:par_var_im_k2}}
\end{figure*}

\begin{figure*}
\centering
\begin{tabular}{cc}
a) $\alpha=60\cdot10^9$ Pa, $\eta_1=8.4\cdot10^{17}$ Pa s, $\eta=9\cdot10^{20}$ Pa s & b) $\alpha_1=61\cdot10^{10}$ Pa, $\eta_1=8.4\cdot10^{17}$ Pa s, $\eta=9\cdot10^{20}$ Pa s\\
\includegraphics[scale=0.46]{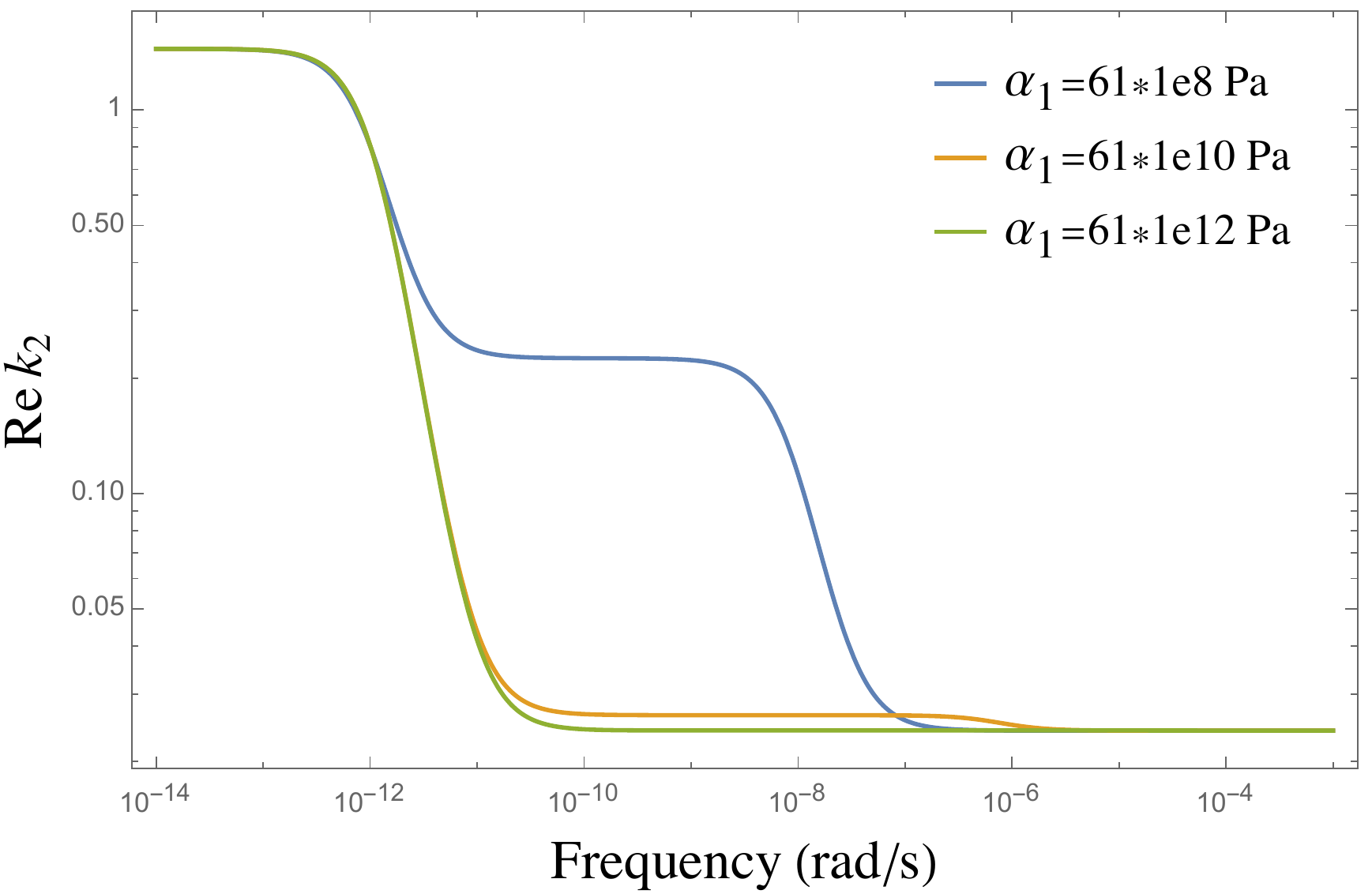} & \includegraphics[scale=0.46]{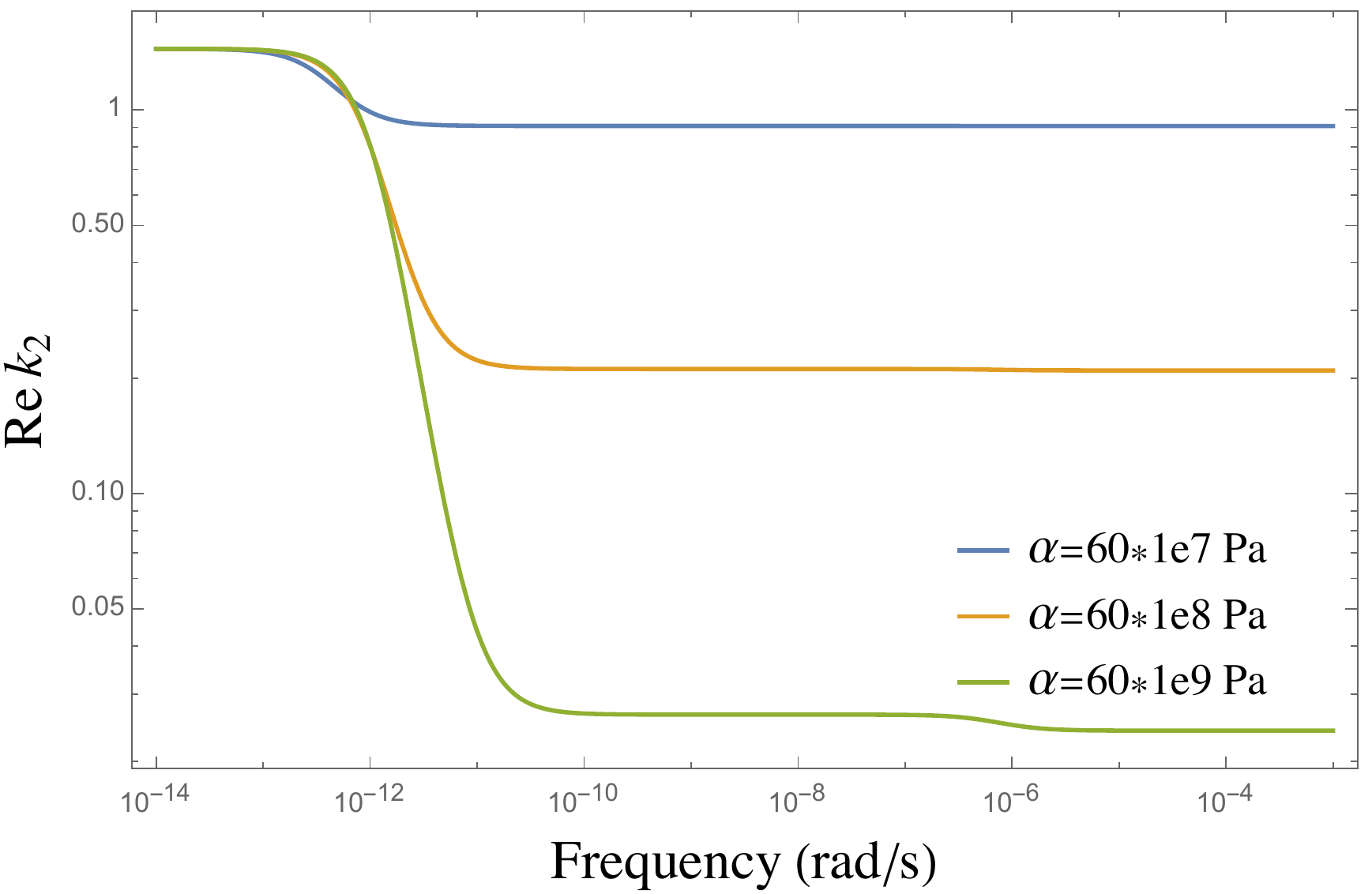}
\vspace{2em}\\
c) $\alpha_1=61\cdot10^{10}$ Pa, $\alpha=60\cdot10^9$ Pa, $\eta=9\cdot10^{20}$ Pa s & d) $\alpha_1=61\cdot10^{10}$ Pa, $\alpha=60\cdot10^9$ Pa, $\eta_1=8.4\cdot10^{17}$ Pa s\\
\includegraphics[scale=0.46]{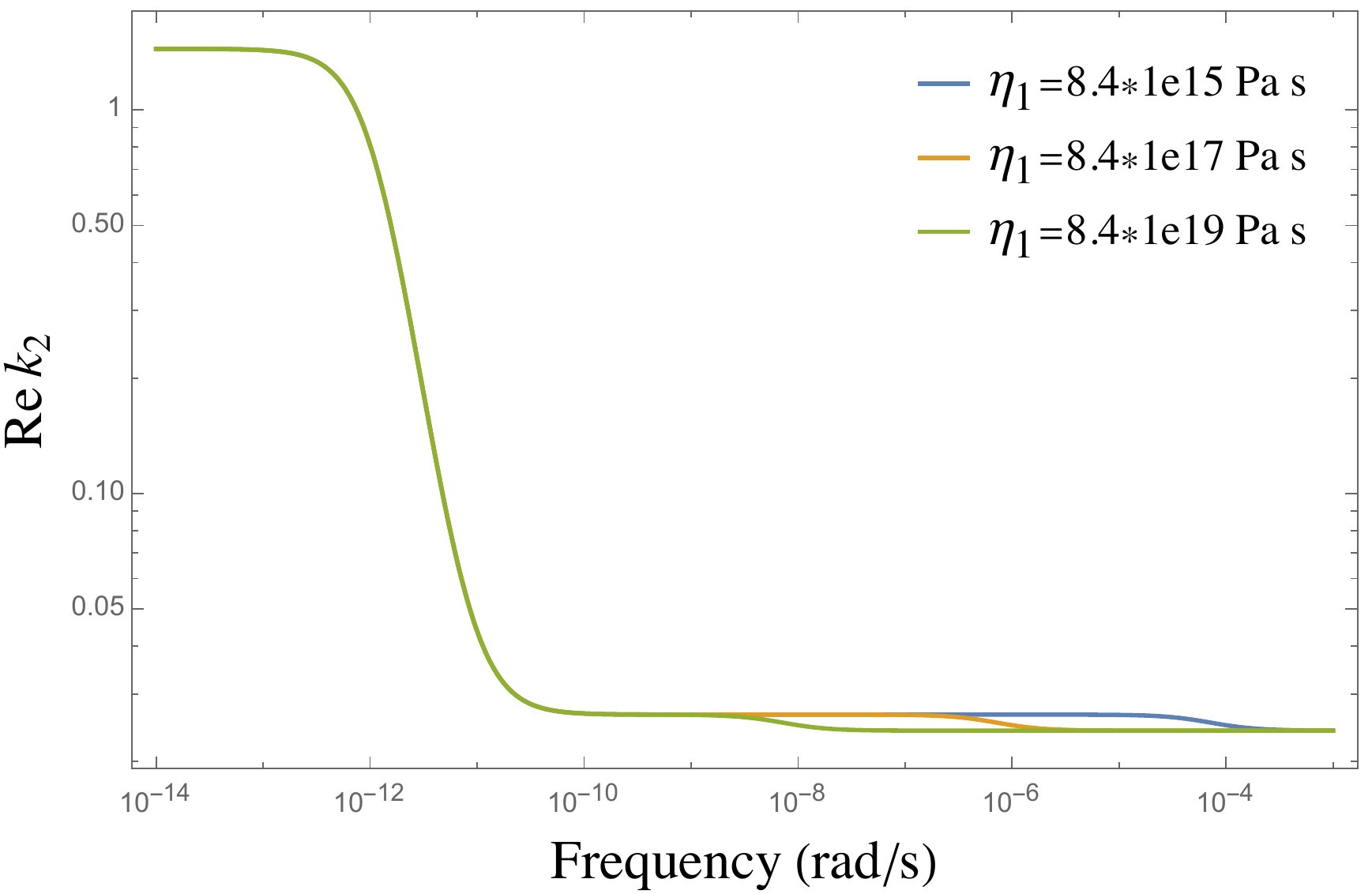} & \includegraphics[scale=0.46]{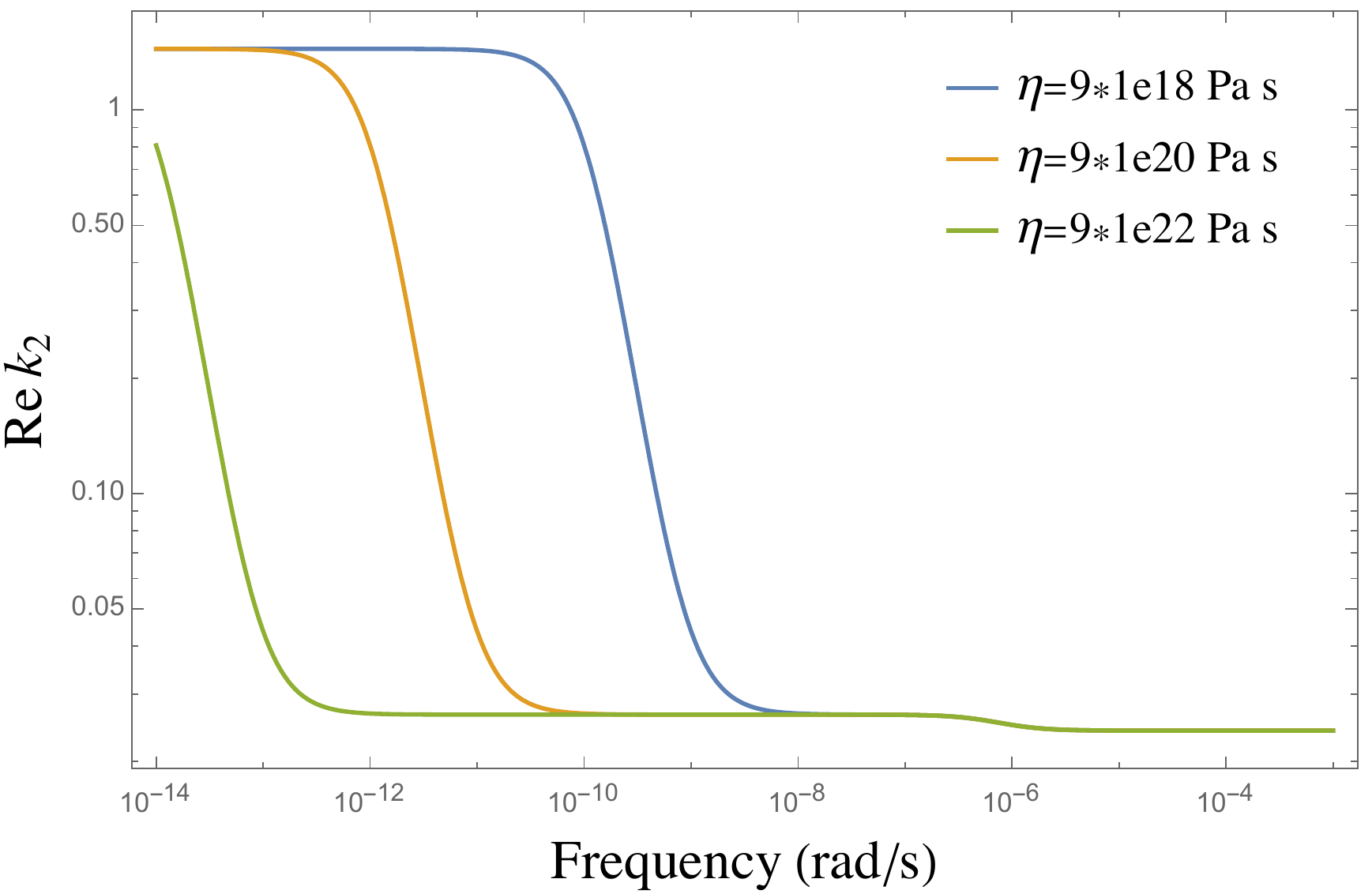}\\
\end{tabular}
\caption{Real part of Love number as a function of frequency for a homogeneous body modeled with Burgers rheology; (a) effect of the Voigt element elasticity $\alpha_1$; (b) effect of the Maxwell element elasticity $\alpha$; (c) effect of the Voigt element viscosity $\eta_1$; (d) effect of the Maxwell element viscosity $\eta$.\label{fig:par_var_re_k2}}
\end{figure*}

We first see how the dissipation of Burgers model depends on the rheology parameters. The complex compliance for Burgers rheology is \citep{bla2016}
\begin{equation}
J(\omega)=\frac{1}{\alpha}+\frac{1}{i\omega\eta}+\frac{1}{\alpha_{1}+i\omega\eta_{1}},
\end{equation}
where $\alpha$, $\alpha_1$ and $\eta$, $\eta_1$ have the dimension of $\rm{s}^{-2}$ and $\rm{s}^{-1}$ respectively (see Footnote \ref{footnote1}). Tidal Love number is given by Equation (\ref{eq:K2(omega)})\textcolor{red}{.} In Figures \ref{fig:par_var_im_k2} and \ref{fig:par_var_re_k2} we illustrate how the imaginary and real parts of tidal Love number change with variation of rheology parameters of Burgers model, hence they can be used to finely tune the parameters of Burgers rheology to mimic the dissipative behavior of stratified Moon or any other body. 

The best fit is obtained for the following parameters of Voigt rheology (we will call this rheology 2nd rheology) $\alpha=60\;\mathrm{GPa}$, $\eta=9.05\times 10^{20}\;\mathrm{Pa\cdot s}$, $\alpha_1=643 \;\mathrm{GPa}$ and $\eta_1=8.8\times 10^{17} \;\mathrm{Pa\cdot s}$, and can be seen in Figure \ref{fig:fit}. The dissipative response assuming the generalized Voigt rheology with two Voigt elements and the parameters obtained in Section \ref{simplification}, and that of the Burgers rheology with the parameters described above are indistinguishable from each other. The choice of simplification method will depend on the problem one is considering, giving preference to the simplest rhelogical model. 

\begin{figure}
\centering
\includegraphics[scale=0.46]{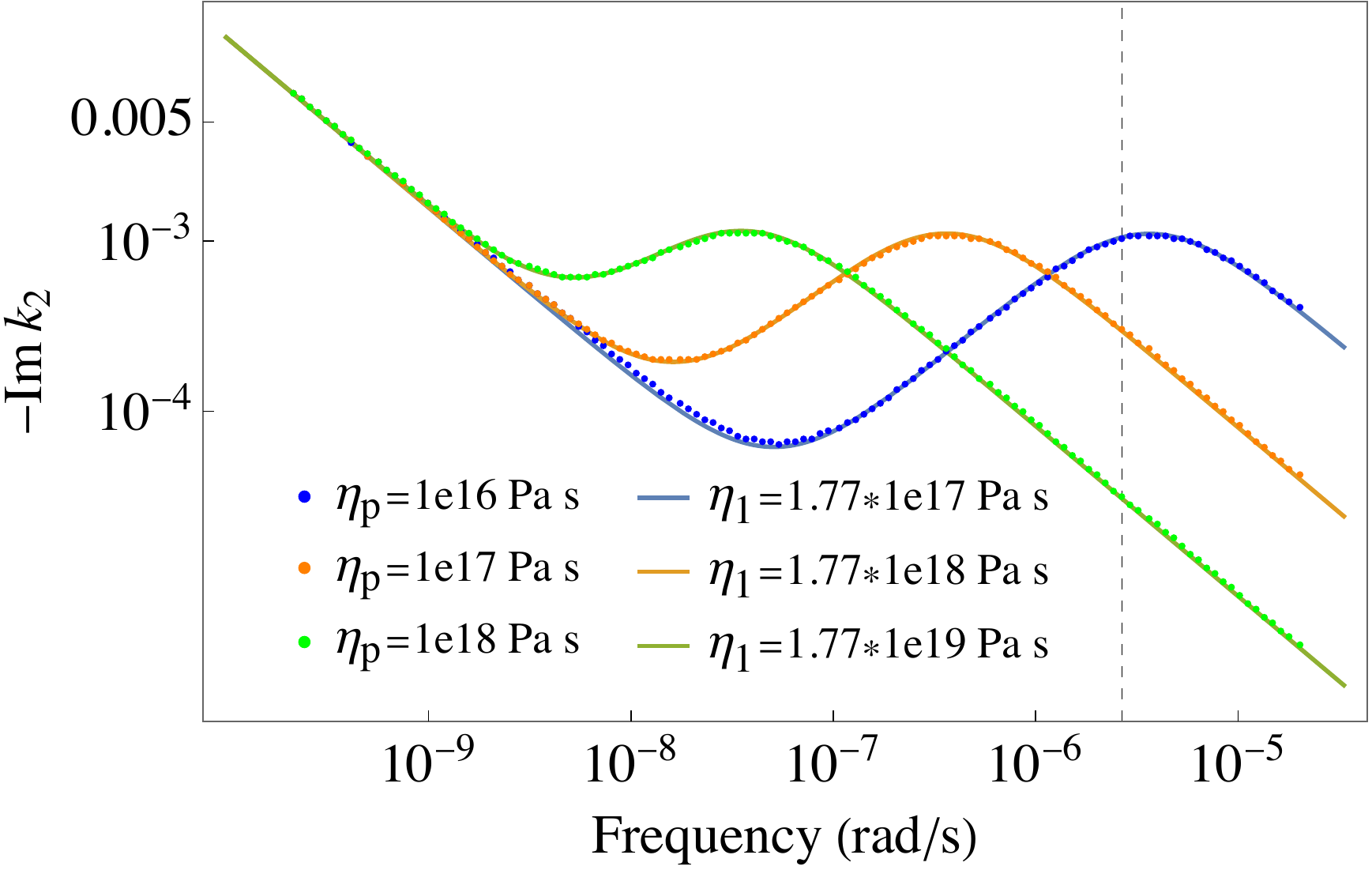}
\caption{Moon dissipation rate variation due to low viscosity layer viscosity variation. Solid lines are obtained by varying the viscosity of low viscosity layer for the stratified Moon. Dotted lines are obtained by varying the viscosity $\eta_1$ of Voigt element in Figure \ref{burgers-osc} or the viscosities $\eta_1$ and $\eta_2$ of Voigt elements of 1st rheology. \label{fig:variation}}
\end{figure}

It would be interesting to establish a straightforward connection between the parameters of a stratified body and that of the corresponding homogeneous rheology. For the stratified Moon the dotted lines in Figure \ref{fig:variation} are obtained varying the viscosity of the low viscosity layer from $10^{16}\;\mathrm{Pa\cdot s}$ to $10^{18}\;\mathrm{Pa\cdot s}$. To mimic the variation in the dissipation rate of the stratified body we vary the viscosities $\eta_1$ and $\eta_2$ of Voigt elements of 1st rheology from $3.32\times 10^{17}\;\mathrm{Pa\cdot s}$ to $3.32\times 10^{19} \;\mathrm{Pa\cdot s}$ and from $3.52\times 10^{17} \;\mathrm{Pa\cdot s}$ to $3.52\times 10^{19} \;\mathrm{Pa\cdot s}$, respectively, or the viscosity $\eta_1$ of Voigt element of 2nd rheology from $1.77\times 10^{17} \;\mathrm{Pa\cdot s}$ to $1.77\times 10^{19} \;\mathrm{Pa\cdot s}$. We then conclude that the change of the viscosity of the low viscosity layer can be reproduced by a change of the same order of magnitude in the viscosity of Voigt elements of the simplified rheologies.


\section{Conclusions}\label{conclusions}

We revisited the long standing problem of approximating the tidal response of a stratified body by that of a homogeneous body. We show that the frequency dependence of tidal dissipation and the quality factor of a multilayered body can be approximated by that of a homogeneous body with complex rheology. This result highlights the fact that we do not need the complexity of the multilayer planet model in order to estimate its tidal dissipation. On the example of the Moon we illustrate that stratified and homogeneous models cannot be distinguished from each other only by the measurement of second degree tidal Love number and quality factor.

We propose and compare two distinct approaches to associate a simple homogeneous rheology to a given stratified moon or planet. The obtained homogeneous rheology can then be used in tandem with the formalism proposed and developed in \citep{RaR2017, Correia2018, gev2020, ragazzo2022librations} to perform a fully three dimensional numerical simulation of the dynamics of a system of many deformable bodies.

It is important to establish a straightforward connection between the parameters of the stratified body and that of the homogeneous rheology. Here we show, on the example of the Moon, that the variation of viscosity of partial melt layer results in the same order variation of the viscosity of the Voigt elements in generalized Voigt model used to approximate the stratified body. We will revisit the problem to establish more general relations between the parameters in future work.

\section*{Acknowledgements}

The authors thank an anonymous referee for several remarks and many suggestions improving the manuscript. YG is partially supported by FAPESP grants 2019/25356-9 and 2021/09679-2. CR is partially supported by FAPESP grant 2016/25053-8.

\section*{Data Availability}

The data underlying this paper will be shared on reasonable request to the corresponding author.
 



\bibliographystyle{mnras}
\bibliography{mybibliography} 

\begin{thebibliography}{}
\makeatletter
\relax
\def\mn@urlcharsother{\let\do\@makeother \do\$\do\&\do\#\do\^\do\_\do\%\do\~}
\def\mn@doi{\begingroup\mn@urlcharsother \@ifnextchar [ {\mn@doi@}
  {\mn@doi@[]}}
\def\mn@doi@[#1]#2{\def\@tempa{#1}\ifx\@tempa\@empty \href
  {http://dx.doi.org/#2} {doi:#2}\else \href {http://dx.doi.org/#2} {#1}\fi
  \endgroup}
\def\mn@eprint#1#2{\mn@eprint@#1:#2::\@nil}
\def\mn@eprint@arXiv#1{\href {http://arxiv.org/abs/#1} {{\tt arXiv:#1}}}
\def\mn@eprint@dblp#1{\href {http://dblp.uni-trier.de/rec/bibtex/#1.xml}
  {dblp:#1}}
\def\mn@eprint@#1:#2:#3:#4\@nil{\def\@tempa {#1}\def\@tempb {#2}\def\@tempc
  {#3}\ifx \@tempc \@empty \let \@tempc \@tempb \let \@tempb \@tempa \fi \ifx
  \@tempb \@empty \def\@tempb {arXiv}\fi \@ifundefined
  {mn@eprint@\@tempb}{\@tempb:\@tempc}{\expandafter \expandafter \csname
  mn@eprint@\@tempb\endcsname \expandafter{\@tempc}}}

\bibitem[\protect\citeauthoryear{{Andrade}}{{Andrade}}{1910}]{Andrade1910}
{Andrade} E.~N. D.~C.,  1910, Proceedings of the Royal Society of London Series
  A, \href {https://ui.adsabs.harvard.edu/abs/1910RSPSA..84....1A} {84, 1}

\bibitem[\protect\citeauthoryear{{Bagheri} et~al.,}{{Bagheri}
  et~al.}{2022}]{Bagheri2022}
{Bagheri} A.,  et~al., 2022, \mn@doi [Advances in Geophysics]
  {10.1016/bs.agph.2022.07.004}, \href
  {https://ui.adsabs.harvard.edu/abs/2022AdGeo..63..231B} {63, 231}

\bibitem[\protect\citeauthoryear{Bland}{Bland}{2016}]{bla2016}
Bland D.~R.,  2016, {The Theory of Linear Viscoelasticity}.
Dover Books on Physics, Dover Publications, \url
  {https://books.google.com.br/books?id=YL8zDQAAQBAJ}

\bibitem[\protect\citeauthoryear{{Bolmont}, {Breton}, {Tobie}, {Dumoulin},
  {Mathis}  \& {Grasset}}{{Bolmont} et~al.}{2020}]{Bolmont2020}
{Bolmont} E.,  {Breton} S.~N.,  {Tobie} G.,  {Dumoulin} C.,  {Mathis} S.,
  {Grasset} O.,  2020, \mn@doi [A\&A] {10.1051/0004-6361/202038204}, \href
  {https://ui.adsabs.harvard.edu/abs/2020A&A...644A.165B} {644, A165}

\bibitem[\protect\citeauthoryear{{Bou{\'e}}, {Rambaux}  \&
  {Richard}}{{Bou{\'e}} et~al.}{2017}]{Boue2017}
{Bou{\'e}} G.,  {Rambaux} N.,   {Richard} A.,  2017, \mn@doi [Celestial
  Mechanics and Dynamical Astronomy] {10.1007/s10569-017-9790-8}, 129, 449

\bibitem[\protect\citeauthoryear{{Carr} et~al.,}{{Carr}
  et~al.}{1998}]{Carr1998}
{Carr} M.~H.,  et~al., 1998, \mn@doi [Nature] {10.1038/34857}, \href
  {https://ui.adsabs.harvard.edu/abs/1998Natur.391..363C} {391, 363}

\bibitem[\protect\citeauthoryear{{Correia}, {Ragazzo}  \& {Ruiz}}{{Correia}
  et~al.}{2018}]{Correia2018}
{Correia} A.~C.~M.,  {Ragazzo} C.,   {Ruiz} L.~S.,  2018, \mn@doi [Celestial
  Mechanics and Dynamical Astronomy] {10.1007/s10569-018-9847-3}, \href
  {https://ui.adsabs.harvard.edu/abs/2018CeMDA.130...51C} {130, 51}

\bibitem[\protect\citeauthoryear{{Dziewonski} \& {Anderson}}{{Dziewonski} \&
  {Anderson}}{1981}]{Dziewonski1981}
{Dziewonski} A.~M.,  {Anderson} D.~L.,  1981, \mn@doi [Physics of the Earth and
  Planetary Interiors] {10.1016/0031-9201(81)90046-7}, \href
  {https://ui.adsabs.harvard.edu/abs/1981PEPI...25..297D} {25, 297}

\bibitem[\protect\citeauthoryear{{Folonier} \& {Ferraz-Mello}}{{Folonier} \&
  {Ferraz-Mello}}{2017}]{Folonier2017}
{Folonier} H.~A.,  {Ferraz-Mello} S.,  2017, \mn@doi [Celestial Mechanics and
  Dynamical Astronomy] {10.1007/s10569-017-9777-5}, 129, 359

\bibitem[\protect\citeauthoryear{Garcia, Gagnepain-Beyneix, Chevrot  \&
  Lognonn\'e}{Garcia et~al.}{2011}]{Garcia.2011}
Garcia R.~F.,  Gagnepain-Beyneix J.,  Chevrot S.,   Lognonn\'e P.,  2011,
  \mn@doi [Physics of the Earth and Planetary Interiors]
  {10.1016/j.pepi.2011.06.015}, 188, 96

\bibitem[\protect\citeauthoryear{{Gevorgyan}}{{Gevorgyan}}{2021}]{Gev2021}
{Gevorgyan} Y.,  2021, \mn@doi [A\&A] {10.1051/0004-6361/202140736}, \href
  {https://ui.adsabs.harvard.edu/abs/2021A&A...650A.141G} {650, A141}

\bibitem[\protect\citeauthoryear{{Gevorgyan}, {Bou{\'e}}, {Ragazzo}, {Ruiz}  \&
  {Correia}}{{Gevorgyan} et~al.}{2020}]{gev2020}
{Gevorgyan} Y.,  {Bou{\'e}} G.,  {Ragazzo} C.,  {Ruiz} L.~S.,   {Correia}
  A.~C.~M.,  2020, \mn@doi [Icarus] {10.1016/j.icarus.2019.113610}, \href
  {https://ui.adsabs.harvard.edu/abs/2020Icar..34313610G} {343, 113610}

\bibitem[\protect\citeauthoryear{{Goossens}, {Renaud}, {Henning}, {Mazarico},
  {Bertone}  \& {Genova}}{{Goossens} et~al.}{2022}]{Goossens2022}
{Goossens} S.,  {Renaud} J.~P.,  {Henning} W.~G.,  {Mazarico} E.,  {Bertone}
  S.,   {Genova} A.,  2022, \mn@doi [The Planetary Science Journal]
  {10.3847/PSJ/ac4bb8}, \href
  {https://ui.adsabs.harvard.edu/abs/2022PSJ.....3...37G} {3, 37}

\bibitem[\protect\citeauthoryear{{Harada}, {Goossens}, {Matsumoto}, {Yan},
  {Ping}, {Noda}  \& {Haruyama}}{{Harada} et~al.}{2014}]{Harada2014}
{Harada} Y.,  {Goossens} S.,  {Matsumoto} K.,  {Yan} J.,  {Ping} J.,  {Noda}
  H.,   {Haruyama} J.,  2014, \mn@doi [Nature Geoscience] {10.1038/ngeo2211},
  \href {https://ui.adsabs.harvard.edu/abs/2014NatGe...7..569H} {7, 569}

\bibitem[\protect\citeauthoryear{{Iess} et~al.,}{{Iess}
  et~al.}{2012}]{iess2012}
{Iess} L.,  et~al., 2012, \mn@doi [Science] {10.1126/science.1219631}, \href
  {https://ui.adsabs.harvard.edu/abs/2012Sci...337..457I} {337, 457}

\bibitem[\protect\citeauthoryear{Jara-Oru{\'e} \& Vermeersen}{Jara-Oru{\'e} \&
  Vermeersen}{2011}]{jara2011effects}
Jara-Oru{\'e} H.~M.,  Vermeersen B.~L.~A.,  2011, Icarus, 215, 417

\bibitem[\protect\citeauthoryear{{Love}}{{Love}}{1911}]{Love1911}
{Love} A.~E.~H.,  1911, {Some Problems of Geodynamics}.
Cambridge University Press

\bibitem[\protect\citeauthoryear{Mathews, Herring  \& Buffett}{Mathews
  et~al.}{2002}]{mathews2002modeling}
Mathews P.~M.,  Herring T.~A.,   Buffett B.~A.,  2002, Journal of Geophysical
  Research: Solid Earth, 107, ETG

\bibitem[\protect\citeauthoryear{Matsumoto, Yamada, Kikuchi, Kamata, Ishihara,
  Iwata, Hanada  \& Sasaki}{Matsumoto et~al.}{2015}]{Matsumoto.2015}
Matsumoto K.,  Yamada R.,  Kikuchi F.,  Kamata S.,  Ishihara Y.,  Iwata T.,
  Hanada H.,   Sasaki S.,  2015, \mn@doi [Geophys. Res. Lett.]
  {10.1002/2015gl065335}, 42, 7351

\bibitem[\protect\citeauthoryear{{Matsuyama}}{{Matsuyama}}{2014}]{Matsuyama2014}
{Matsuyama} I.,  2014, \mn@doi [Icarus] {10.1016/j.icarus.2014.07.005}, \href
  {https://ui.adsabs.harvard.edu/abs/2014Icar..242...11M} {242, 11}

\bibitem[\protect\citeauthoryear{{Matsuyama}, {Nimmo}, {Keane}, {Chan},
  {Taylor}, {Wieczorek}, {Kiefer}  \& {Williams}}{{Matsuyama}
  et~al.}{2016}]{Matsuyama2016}
{Matsuyama} I.,  {Nimmo} F.,  {Keane} J.~T.,  {Chan} N.~H.,  {Taylor} G.~J.,
  {Wieczorek} M.~A.,  {Kiefer} W.~S.,   {Williams} J.~G.,  2016, \mn@doi
  [geophysical Research Letters] {10.1002/2016GL069952}, \href
  {https://ui.adsabs.harvard.edu/abs/2016GeoRL..43.8365M} {43, 8365}

\bibitem[\protect\citeauthoryear{{Matsuyama}, {Beuthe}, {Hay}, {Nimmo}  \&
  {Kamata}}{{Matsuyama} et~al.}{2018}]{Matsuyama2018}
{Matsuyama} I.,  {Beuthe} M.,  {Hay} H.~C.~F.~C.,  {Nimmo} F.,   {Kamata} S.,
  2018, \mn@doi [Icarus] {10.1016/j.icarus.2018.04.013}, \href
  {https://ui.adsabs.harvard.edu/abs/2018Icar..312..208M} {312, 208}

\bibitem[\protect\citeauthoryear{{Mayor} \& {Queloz}}{{Mayor} \&
  {Queloz}}{1995}]{Mayor1995}
{Mayor} M.,  {Queloz} D.,  1995, \mn@doi [Nature] {10.1038/378355a0}, \href
  {https://ui.adsabs.harvard.edu/abs/1995Natur.378..355M} {378, 355}

\bibitem[\protect\citeauthoryear{{Nimmo}, {Faul}  \& {Garnero}}{{Nimmo}
  et~al.}{2012}]{Nimmo2012}
{Nimmo} F.,  {Faul} U.~H.,   {Garnero} E.~J.,  2012, \mn@doi [Journal of
  Geophysical Research (Planets)] {10.1029/2012JE004160}, \href
  {https://ui.adsabs.harvard.edu/abs/2012JGRE..117.9005N} {117, E09005}

\bibitem[\protect\citeauthoryear{Poirier, Boloh  \& Chambon}{Poirier
  et~al.}{1983}]{Poirier.1983}
Poirier J.~P.,  Boloh L.,   Chambon P.,  1983, \mn@doi [Icarus]
  {10.1016/0019-1035(83)90076-3}, 55, 218

\bibitem[\protect\citeauthoryear{Ragazzo}{Ragazzo}{2020}]{ragazzo2020theory}
Ragazzo C.,  2020, S{\~a}o Paulo Journal of Mathematical Sciences, 14, 1

\bibitem[\protect\citeauthoryear{{Ragazzo} \& {Ruiz}}{{Ragazzo} \&
  {Ruiz}}{2017}]{RaR2017}
{Ragazzo} C.,  {Ruiz} L.~S.,  2017, \mn@doi [Celestial Mechanics and Dynamical
  Astronomy] {10.1007/s10569-016-9741-9}, \href
  {https://ui.adsabs.harvard.edu/abs/2017CeMDA.128...19R} {128, 19}

\bibitem[\protect\citeauthoryear{{Ragazzo}, {Bou{\'e}}, {Gevorgyan}  \&
  {Ruiz}}{{Ragazzo} et~al.}{2022}]{ragazzo2022librations}
{Ragazzo} C.,  {Bou{\'e}} G.,  {Gevorgyan} Y.,   {Ruiz} L.~S.,  2022, \mn@doi
  [Celestial Mechanics and Dynamical Astronomy] {10.1007/s10569-021-10055-3},
  \href {https://ui.adsabs.harvard.edu/abs/2022CeMDA.134...10R} {134, 10}

\bibitem[\protect\citeauthoryear{Sabadini, Vermeersen  \& Cambiotti}{Sabadini
  et~al.}{2016}]{sabadini2016global}
Sabadini R.,  Vermeersen B.,   Cambiotti G.,  2016, Global dynamics of the
  Earth.
Springer

\bibitem[\protect\citeauthoryear{{Segatz}, {Spohn}, {Ross}  \&
  {Schubert}}{{Segatz} et~al.}{1988}]{Segatz1988}
{Segatz} M.,  {Spohn} T.,  {Ross} M.~N.,   {Schubert} G.,  1988, \mn@doi
  [Icarus] {10.1016/0019-1035(88)90001-2}, \href
  {https://ui.adsabs.harvard.edu/abs/1988Icar...75..187S} {75, 187}

\bibitem[\protect\citeauthoryear{{Shoji}, {Hussmann}, {Kurita}  \&
  {Sohl}}{{Shoji} et~al.}{2013}]{Shoji2013}
{Shoji} D.,  {Hussmann} H.,  {Kurita} K.,   {Sohl} F.,  2013, \mn@doi [Icarus]
  {10.1016/j.icarus.2013.05.004}, \href
  {https://ui.adsabs.harvard.edu/abs/2013Icar..226...10S} {226, 10}

\bibitem[\protect\citeauthoryear{{Sundberg} \& {Cooper}}{{Sundberg} \&
  {Cooper}}{2010}]{Sundberg2010}
{Sundberg} M.,  {Cooper} R.~F.,  2010, \mn@doi [Philosophical Magazine]
  {10.1080/14786431003746656}, \href
  {https://ui.adsabs.harvard.edu/abs/2010PMag...90.2817S} {90, 2817}

\bibitem[\protect\citeauthoryear{Tanaka, Okuno  \& Okubo}{Tanaka
  et~al.}{2006}]{tanaka2006new}
Tanaka Y.,  Okuno J.,   Okubo S.,  2006, Geophysical Journal International,
  164, 273

\bibitem[\protect\citeauthoryear{{Thomas}, {Tajeddine}, {Tiscareno}, {Burns},
  {Joseph}, {Loredo}, {Helfenstein}  \& {Porco}}{{Thomas}
  et~al.}{2016}]{THOMAS201637}
{Thomas} P.~C.,  {Tajeddine} R.,  {Tiscareno} M.~S.,  {Burns} J.~A.,  {Joseph}
  J.,  {Loredo} T.~J.,  {Helfenstein} P.,   {Porco} C.,  2016, \mn@doi [Icarus]
  {10.1016/j.icarus.2015.08.037}, \href
  {https://ui.adsabs.harvard.edu/abs/2016Icar..264...37T} {264, 37}

\bibitem[\protect\citeauthoryear{{Walterov{\'a}} \&
  {B{\v{e}}hounkov{\'a}}}{{Walterov{\'a}} \&
  {B{\v{e}}hounkov{\'a}}}{2017}]{Walterova2017}
{Walterov{\'a}} M.,  {B{\v{e}}hounkov{\'a}} M.,  2017, \mn@doi [Celestial
  Mechanics and Dynamical Astronomy] {10.1007/s10569-017-9772-x}, \href
  {https://ui.adsabs.harvard.edu/abs/2017CeMDA.129..235W} {129, 235}

\bibitem[\protect\citeauthoryear{{Walterov{\'a}}, {B{\v{e}}hounkov{\'a}}  \&
  {Efroimsky}}{{Walterov{\'a}} et~al.}{2023}]{Walterova2023}
{Walterov{\'a}} M.,  {B{\v{e}}hounkov{\'a}} M.,   {Efroimsky} M.,  2023,
  \mn@doi [arXiv e-prints] {10.48550/arXiv.2301.02476}, \href
  {https://ui.adsabs.harvard.edu/abs/2023arXiv230102476W} {p. arXiv:2301.02476}

\bibitem[\protect\citeauthoryear{Weber, Lin, Garnero, Williams  \&
  Lognonné}{Weber et~al.}{2011}]{Weber.2011}
Weber R.~C.,  Lin P.~Y.,  Garnero E.~J.,  Williams Q.,   Lognonné P.,  2011,
  \mn@doi [Science] {10.1126/science.1199375}, 331, 309

\bibitem[\protect\citeauthoryear{{Williams} \& {Boggs}}{{Williams} \&
  {Boggs}}{2015}]{Williams2015}
{Williams} J.~G.,  {Boggs} D.~H.,  2015, \mn@doi [Journal of Geophysical
  Research (Planets)] {10.1002/2014JE004755}, \href
  {https://ui.adsabs.harvard.edu/abs/2015JGRE..120..689W} {120, 689}

\bibitem[\protect\citeauthoryear{Wu \& Peltier}{Wu \&
  Peltier}{1982}]{wu1982viscous}
Wu P.,  Peltier W.~R.,  1982, Geophysical Journal International, 70, 435

\makeatother
\end{thebibliography}




\appendix

\section{Dominant modes in the normal mode solution for the stratified Moon}\label{modes}

\begin{figure}
\centering
\includegraphics[scale=0.46]{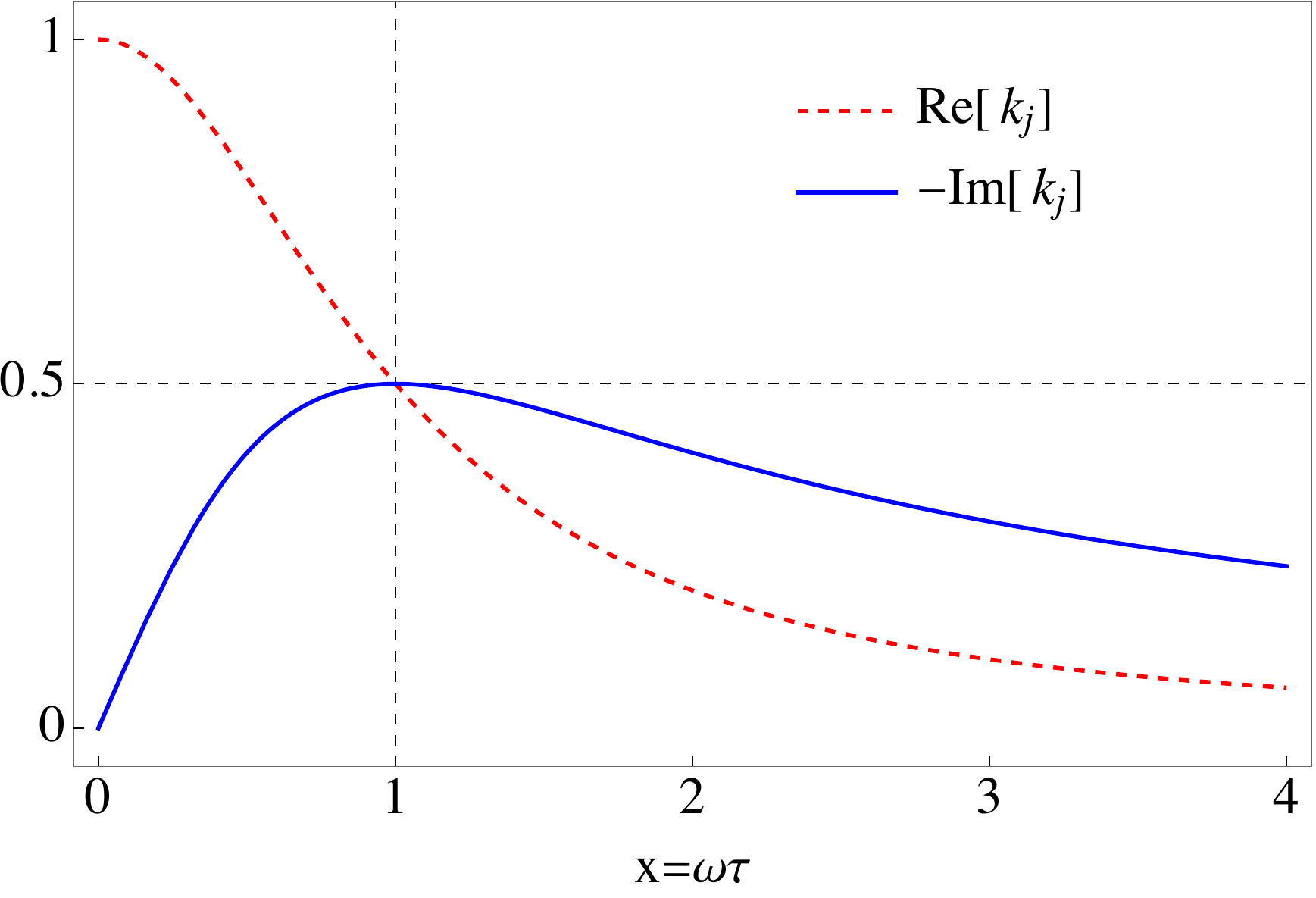}\\
\caption{The graphs of the real (red-dashed) and the imaginary (blue) parts of a term (mode) in the partial 
fraction expansion of $k_2(i\omega)$. 
The amplitude $r_j\tau_j$ was normalized to one and  the angular frequency
$x=\omega \tau_j$ was normalized by the relaxation time $\tau_j$ of the mode. The maximum of the imaginary part, the point of maximum dissipation
of energy, occurs at $x=1$, which is also the half-width of the real part. \label{term}}
\end{figure}

The tidal force may be decomposed into harmonic components. Let $\omega>0$ be the angular frequency of one of these components. The tidal response of the body at frequency $\omega$ is determined by $k(i\omega)$. Each term in the expansion (\ref{saba367}) has real and imaginary parts given, respectively, by
\begin{equation}
\Re[k_{j}]=\tau_j r_j\frac{1}{1+x^2},\quad \Im[k_{j}]=-\tau_j r_j\frac{x}{1+x^2},\quad\text{where}\quad x=\omega\tau_j.
\end{equation}
The graphs of $\Re[k_{j}]$ and $\Im[k_{j}]$ are shown in Figure \ref{term}.

Since the maximum  of the imaginary part of a mode $j$ is at $\omega=1/\tau_j$, different modes have their maxima at different points. The modes with larger normalized amplitudes $r_j \tau_j$ are dominant at frequency $\omega=1/\tau_j$.

If two modes $s$ and $g$ satisfy $r_s\tau_s/(r_g \tau_g)< 1 $, then the relative importance of the imaginary part of the modes at a point 
of maximum $\omega=1/\tau_s$ of the smaller mode is measured by the ratio 
\begin{equation}\Im[k_{s}](\tau_s^{-1})/\Im[k_{g}](\tau_s^{-1})=\frac{r_s\tau_s}{2r_g \tau_g}\frac{1+(\tau_g/\tau_s)^2}{\tau_g/\tau_s}\label{kI}.\end{equation}

\begin{figure}
\centering
\includegraphics[scale=0.46]{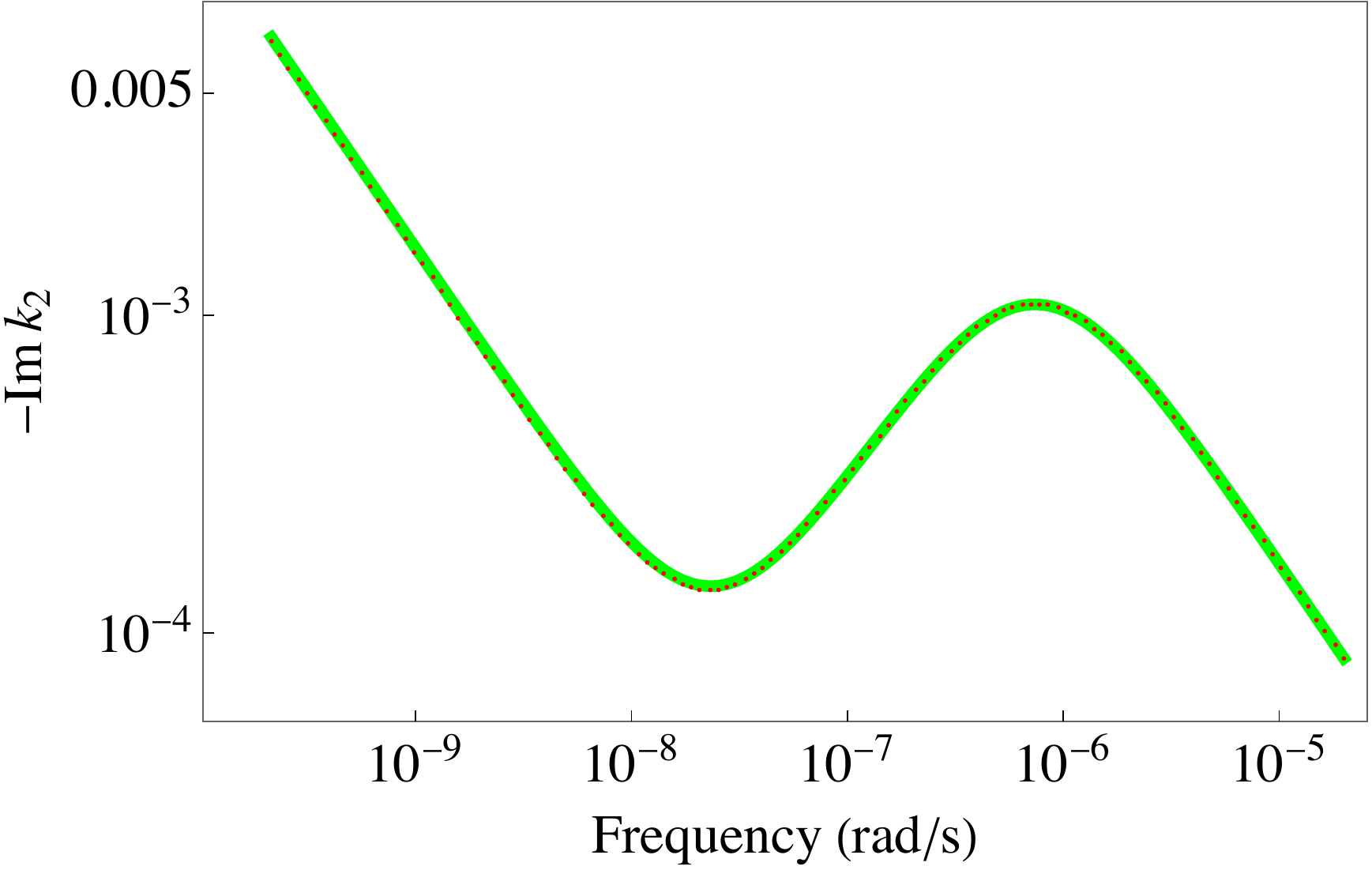}\\
\vspace{2em}
\includegraphics[scale=0.46]{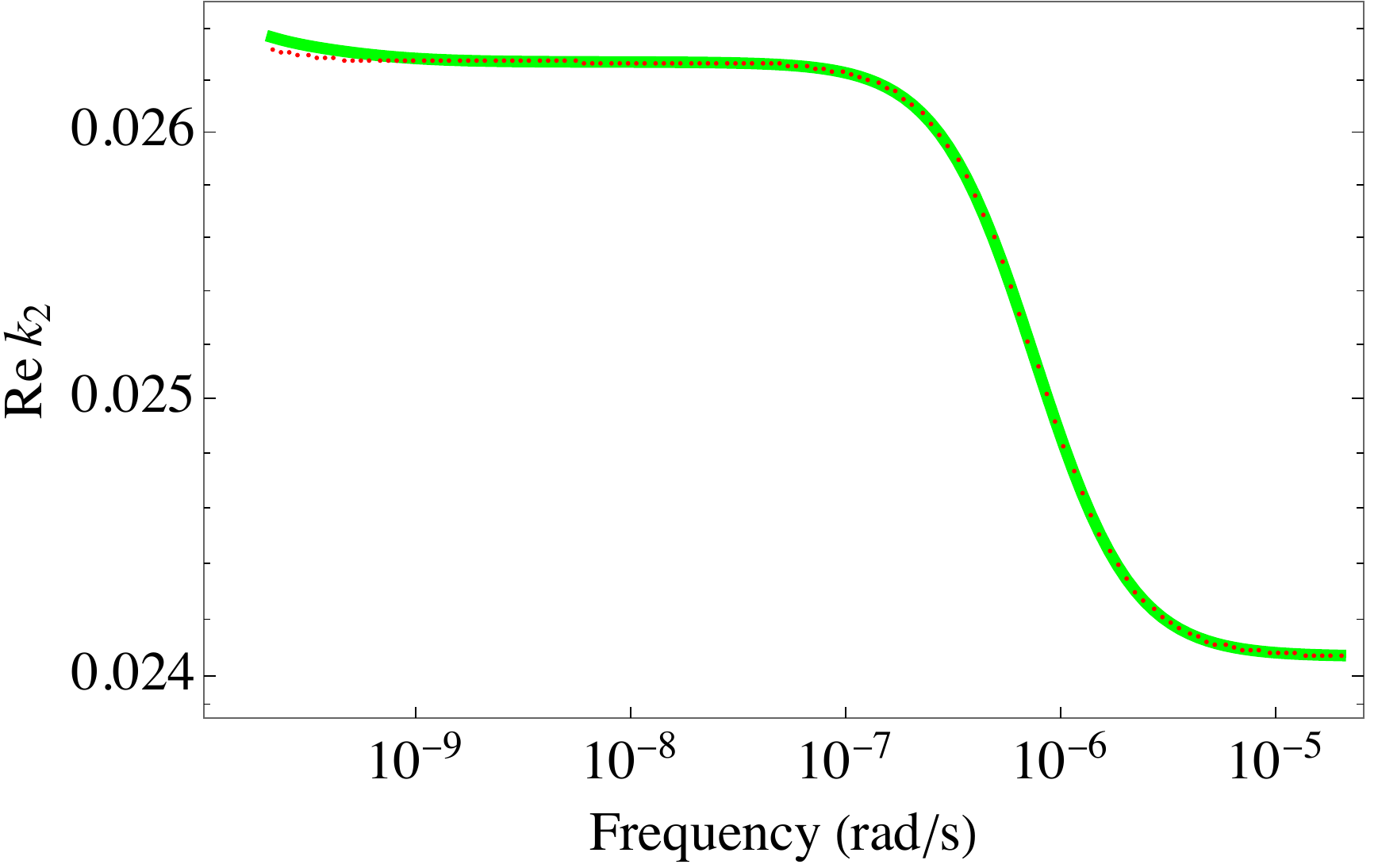}
\caption{Tidal Love number dependence on frequency for the Moon. The green solid line was plotted using all the terms and the red dotted line was plotted using only the terms highlighted in red terms in Table \ref{k2_expansion}. \label{fig:k2_exp}}
\end{figure}

For the Moon, in the interval from $2\times10^{-5}$ to $2\times10^{-10}$ rad/s,  most relevant modes are 5, 6 and 9 (see Table \ref{k2_expansion}) and $k_2(i\omega)$ is mostly determined by the modes 5, 6 and 9 and by the elastic Love number $k_E$ (see Figure \ref{fig:k2_exp}). 

If two modes $s$ and $g$ satisfy $r_s\tau_s/(r_g \tau_g)\ll 1$ and $\tau_s>\tau_g$, then $r_s\tau_s/(r_g \tau_g)=\Re[k_{s}](0)/\Re[k_{g}](0)\ge \Re[k_{s}](\omega)/\Re[k_{g}](\omega)$ for all $\omega\ge 0$, and the real part of mode $s$ can be safely neglected.
If $r_s\tau_s/(r_g \tau_g)\ll 1$ but $\tau_s<\tau_g$, then $\Re[k_{s}](\omega)/\Re[k_{g}](\omega)$ is an increasing function of $\omega$ with its maximum at $\omega=\infty$ given by $\frac{r_s\tau_g}{r_g\tau_s}$. The smaller mode $s$ might have some importance at high frequencies.

To illustrate the discussion in the previous paragraphs, we will analyze the effect of the mode 7 in Table \ref{k2_expansion} on the plots of the real and the imaginary parts of $k_2(i\omega)$ shown in Figure \ref{fig:k2_exp}. The mode 7 has a normalized amplitude $r_7\tau_7=0.000057$ and a relaxation time $\tau_7=2\times10^9$ seconds. Compared to the dominant mode 9: $r_7\tau_7/(r_9 \tau_9)=0.000054$ and 
$\tau_7/\tau_9=1/357$. At the frequency $\omega=1/\tau_7$, where $\Im[k_{7}](\omega)$ is maximum, the ratio in equation (\ref{kI}) is $\Im[k_{7}](\tau_7^{-1})/\Im[k_{9}](\tau_7^{-1})=0.01$. The effect of the mode 7 at frequency $5\times 10^{-10}$(rad/sec) is, therefore,  100 times smaller than the effect of the mode 9. Compared to the dominant mode 6: $r_7\tau_7/(r_6 \tau_6)=0.042$ and $\tau_7/\tau_6=1268.4$, consequently the effect of the mode 7 is at least $25$ times smaller than the effect of the mode 6 on the real part of $k_2(i\omega)$. This explains why mode 7 can be neglected.

\bsp	
\label{lastpage}
\end{document}